\documentclass[aoas,nameyear,dvips]{arximspdf}
\usepackage{graphicx}

\doi{10.1214/10-AOAS429}
\volume{5}
\issue{2B}
\pubyear{2011}
\firstpage{1262}
\lastpage{1292}

\makeatletter
\newtheorem{pn}{Proposition}
\newtheorem{tm}{Theorem}
\newtheorem{cy}{Corollary}
\makeatother

\begin{document}
\begin{frontmatter}

\title{Assessment of synchrony in multiple neural spike trains using loglinear point process models}
\runtitle{Loglinear models of synchrony}

\begin{aug}
\author[A]{\fnms{Robert E.} \snm{Kass}\corref{}\ead[label=e1]{kass@stat.cmu.edu}},
\author[A]{\fnms{Ryan C.} \snm{Kelly}\ead[label=e2]{ryekelly@gmail.com}}
\and
\author[B]{\fnms{Wei-Liem} \snm{Loh}\ead[label=e3]{stalohwl@nus.edu.sg}}
\runauthor{R. E. Kass, R. C. Kelly and W.-L. Loh }
\affiliation{Carnegie Mellon University, Carnegie Mellon University\break and National University of Singapore}
\address[A]{R. E. Kass\\
R. C. Kelly\\
Department of Statistics\\
Carnegie Mellon University\\
Pittsburgh, Pennsylvania 15213\\
USA\\
\printead{e1}\\
\phantom{E-mail: }\printead*{e2}} 
\address[B]{W.-L. Loh\\
Department of Statistics and \\
\quad Applied Probability\\
National University of Singapore\\
Singapore 117546\\
Republic of Singapore\\
\printead{e3}}
\end{aug}

\received{\smonth{8} \syear{2008}}

\begin{abstract}
Neural spike trains, which are sequences of very brief jumps in voltage
across the cell membrane, were one of the motivating applications for
the development of point process methodology.  Early work required the
assumption of stationarity, but contemporary experiments often use
time-varying stimuli and produce time-varying neural responses. More
recently, many statistical methods have been developed for
nonstationary neural point process data. There has also been much
interest in identifying synchrony, meaning events across two or more
neurons that are nearly simultaneous at the time scale of the
recordings. A natural statistical approach is to discretize time, using
short time bins, and to introduce loglinear models for dependency among
neurons, but previous use of loglinear modeling technology has assumed
stationarity. We introduce a succinct yet powerful class of
time-varying loglinear models by (a) allowing individual-neuron effects
(main effects) to involve time-varying intensities; (b)  also allowing
the individual-neuron effects to involve autocovariation effects
(history effects) due to past spiking, (c) assuming excess synchrony
effects (interaction effects) do not depend on history, and (d)~assuming
all effects vary smoothly across time. Using data from the
primary visual cortex of an anesthetized monkey, we give two examples
in which the rate of synchronous spiking cannot be explained by
stimulus-related changes in individual-neuron effects. In one example,
the excess synchrony disappears when slow-wave ``up'' states are taken
into account as history effects, while in the second example it does
not. Standard point process theory explicitly rules out synchronous
events. To justify our use of continuous-time methodology, we introduce
a framework that incorporates synchronous events and provides
continuous-time loglinear point process approximations to discrete-time
loglinear models.
\end{abstract}


\end{frontmatter}

\section{Introduction}\label{sec1}

One of the most important techniques in learning about the functioning
of the brain has involved examining neuronal activity in laboratory
animals under varying experimental conditions. Neural information is
represented and communicated through series of action potentials, or
spike trains, and the central scientific issue in many studies concerns
the physiological significance that should be attached to a particular
neuron firing pattern in a particular part of the brain. In
addition, a major relatively new effort in neurophysiology involves the
use of multielectrode recording, in which responses from dozens of
neurons are recorded simultaneously. Much current research focuses on
the information that may be contained in the interactions among
neurons. Of particular interest are spiking events that occur across
neurons in close temporal proximity, within or near the typical one
millisecond accuracy of the recording devices.  In this paper we
provide a point process framework for analyzing such nearly synchronous
events.

The use of point processes to describe and analyze spike train data has
been one of the major contributions of statistics to neuroscience. On
the one hand, the observation that individual point processes may be
considered, approximately, to be binary time series allows methods
associated with generalized linear models to be applied [cf. Brillinger
(\citeyear{Brillinger1988}, \citeyear{Brillinger1992})]. On the other hand, basic point process methodology
coming from the continuous-time representation is important both
conceptually and in deriving data-analytic techniques [e.g., the
time-rescaling theorem may be used for goodness of fit and efficient
spike train simulation; see Brown et al. (\citeyear{BBVK2001})]. The ability to go
back and forth between continuous time, where neuroscience and
statistical theory reside, and discrete time, where measurements are
made and data are analyzed, is central to statistical analysis of spike
trains. From the discrete-time perspective, when multiple spike trains
are considered simultaneously it becomes natural to introduce loglinear
models [cf. Martignon et al. (\citeyear{MDLDFV2000})] and a widely read and hotly debated report by
Schneidman et al. (\citeyear{SBSB2006}) examined the extent to which pairwise
dependence among neurons can capture stimulus-related information. A
fundamental limitation of much of the work in this direction, however,
is its reliance on stationarity. The main purpose of the framework
described below is to handle the nonstationarity inherent in
stimulus-response experiments by introducing appropriate loglinear
models while also allowing passage to a continuous-time limit. The
methods laid out here are in the spirit of Ventura, Cai and Kass
(\citeyear{VCK2005}), who proposed a~bootstrap test of time-varying synchrony, but
our methods are different in detail and our framework is much more
general.

\begin{figure}

\includegraphics{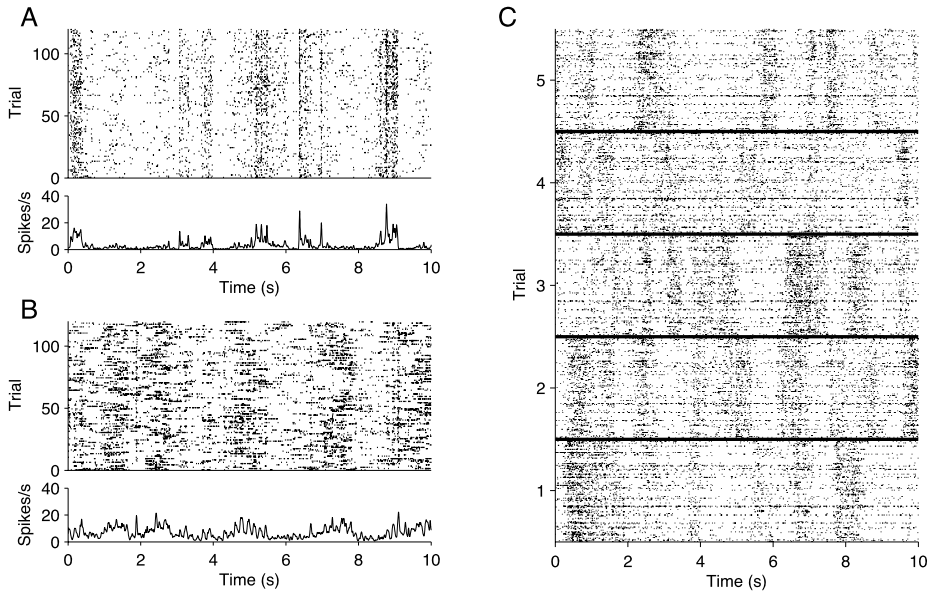}
 \caption{Neural spike train raster plots for repeated presentations of a
drifting sine wave grating stimulus.  \textup{(A)} Single cell responses to 120
repeats of a 10 second movie.  At the top is a raster corresponding to
the spike times, and below is a peri-stimulus time histogram (PSTH) for
the same data.  Portions of the stimulus eliciting firing are apparent.
\textup{(B)} The same plots as in \textup{(A)}, for a different cell.  \textup{(C)} Population
responses to the same stimulus, for 5 repeats.  Each block,
corresponding to a single trial, is the population raster for $\nu=128$
units.  On each trial there are several dark bands, which constitute
bursts of network activity sometimes called ``up states.'' Up state
epochs vary across trials, indicating they are not locked to the
stimulus.}\label{fig1}
\end{figure}

Statistical modeling of point process data focuses on intensity
functions, which represent the rate at which the events occur, and
often involve covariates [cf. Brown et al. (\citeyear{BBEF2003}), Kass, Ventura and
Brown (\citeyear{KVB2005}), Paninski et al. (\citeyear{PBIK2009}) and references therein]. A basic
distinction is that of \textit{conditional} versus \textit{marginal}
intensities: the conditional intensity determines the event rate for a
given realization of the process, while the marginal intensity is the
expectation of the conditional intensity across realizations. In
neurophysiological experiments stimuli are often presented repeatedly
across many trials, resulting in many replications of the multiple
sequences of spike trains. This is the situation we concern ourselves
with here, and it is illustrated in Figure \ref{fig1}, part A, where
the responses of a single neuron for 120 trials are displayed: each
line of the raster plot shows a single spike train, which is the neural
response on a single trial. The experiment that generated these data is
described in Section \ref{sec:example}. The bottom panel in part A of
Figure~\ref{fig1} displays a smoothed peristimulus time histogram
(PSTH), which summarizes the trial-averaged response by pooling across
trials. As we explain in greater detail in Section \ref{sec:overview},
scientific questions and statistical analyses may concern either
within-trial responses (conditional intensities) or trial-averaged
responses (marginal intensities).

A point process evolves in continuous time but, as we have noted, it is
convenient for many statistical purposes to consider a discretized
version. Decomposing time into bins of width $\delta$, we may define a
binary time series to be 1 for every time bin in which an event occurs,
and 0 for every bin in which an event does not occur. It is not hard to
show that, under the usual regularity condition that events occur
discretely (i.e., no two events occur at the same time), the likelihood
function of the binary time series approximates the likelihood of the
point process as $\delta \rightarrow 0$. For a pair of point processes,
the discretized process is a time series of $2\times 2$ polytomous
variables indicating, in each time bin, whether an event of the first
process occurred, an event of the second process occurred, or both, or
neither.  This suggests analyzing nearly synchronous events based on a
loglinear model with cell probabilities that vary across time.
Intuitive as such procedures may be, their point process justification
is subtle: the standard regularity condition forbids two processes
having synchronous events, so it is not obvious how we might obtain
convergence to a point process (as $\delta \rightarrow 0$) for
discrete-process likelihoods that incorporate synchrony.

One way out of this impasse is to introduce a marked point process
framework in which each event/mark could be of three distinct types:
first process, second process, or both.  The standard marked point
process requires modification, however, because it fails to accommodate
independence as a special case. Under independence, the discretized
events for each process occur with probability of order $O(\delta)$,
while the synchronous events occur with probability of order
$O(\delta^2)$ as $\delta \rightarrow 0$. We refer to this as a
\textit{sparsity} condition, and the generalization to multiple processes
involves a \textit{hierarchical sparsity} condition.  Once we introduce a
family of marked point processes indexed by $\delta$, we can guarantee
hierarchical sparsity. Not only does this allow, as it must, the
special case of independence models, but it also makes the conditional
intensity for neuron $i$ depend only on the history for neuron $i$,
asymptotically (as $\delta \rightarrow 0$). This in turn avoids
confounding the dependence described by the loglinear model and greatly
reduces the dimensionality of the problem.  We require two very natural
regularity conditions based on well-known neurophysiology: the
existence of a refractory period, during which the neuron cannot spike
again, and smoothness of the conditional intensity across time. It
would be possible, and sometimes advantageous, instead to model
dependence through the individual-neuron conditional intensity
functions. The loglinear modeling approach used here avoids this
step.

\subsection{A motivating example}\label{sec:example}

In a series of experiments performed by one of us (Kelly, together with
Dr. Matthew Smith), visual images were displayed at resolution $1024
\times 768$ pixels on a computer monitor, while the neural responses in
the primary visual cortex of an anesthetized monkey were recorded. Each
of 98 distinct images consisted of a sinusoidal grating that drifted in
a particular direction for 300 milliseconds, and each was repeated 120
times. Each repetition of the complete sequence of stimuli lasted
approximately 30 seconds. This kind of stimulus has been known to drive
cells in the primary visual cortex since the Nobel prize-winning work
of Hubel and Wiesel in the 1960s. With improved technology and advanced
analytical strategies, much more precise descriptions of neural
response are now possible. A small portion of the data from 5
repetitions of many stimuli is shown in part C of  Figure \ref{fig1}.

The details of the experiment and recording technique are reported in
Kelly et al.~(\citeyear{KSSKBML2007}). A total of 125 neural ``units'' were obtained,
which included about 60 well-isolated individual neurons; the remainder
were of undetermined origin (some mix of 1 or more neurons). The goal
was to discover the interactions among these units in response to the
stimuli. Each neuron will have its own consistent pattern of responses
to stimuli, as illustrated in parts A and B of Figure \ref{fig1}.
Synchronous spiking across neurons is relatively rare. However, in each
of the 5 blocks within part C of Figure \ref{fig1} (each block
corresponding to a single trial) several dark bands of activity across
most neurons may be seen during the trial. These bands correspond to
what are often called network ``up'' states, and are often seen under
anesthesia. For discussion and references see Kelly et al. (\citeyear{KSKL2009}). It
would be of interest to separate the effects of such network activity
from other synchronous activity, especially stimulus-related
synchronous activity. The framework in this paper provides a foundation
for statistical methods that can solve such problems.

\subsection{Overview of approach}\label{sec:overview}

We begin with some notation.  Suppose we observe the activity of an
ensemble of $\nu$ neurons labeled $1$ to $\nu$ over a time interval
$[0,T)$, where $T>0$ is a constant. Let $N_T^i$ denote the total number
of spikes produced by neuron $i$ on $[0,T)$ where $i=1,\ldots, \nu$.
The resulting (stochastic) sequence of spike times is written as $0\leq
s_1^i < \cdots < s^i_{ N^i_T} < T$. For the moment we focus on the case
$\nu=3$, but other values of $\nu$ are of interest and with
contemporary recording technology $\nu \approx 100$ is not uncommon, as
in the experiment in Section~\ref{sec:example}. Let $\delta>0$ be a
constant such that $T$ is a multiple of $\delta$ (for simplicity).  We
divide the time interval into bins of width $\delta$. Define $X^i( t)
=1$ if neuron $i$ has a spike in the time bin $[ t, t + \delta)$ and
$0$ otherwise.  Because of the existence of a~refractory period for
each neuron,
 there can be at most 1 spike in
$[ t, t + \delta)$ from the same neuron if $\delta$ is sufficiently
small.
 Then writing
\[
P^{1, 2, 3}_{a, b, c} (t) = P\bigl( X^1(t)=a, X^2(t) = b, X^3(t) =c\bigr)\qquad\forall a,b,c\in \{0,1\},
\]
the data would involve spike counts across trials [e.g., the number of
trials on which $(X^1(t),$ $X^2(t), X^3(t)) = (1,1,1)$]. The obvious
statistical tool for analyzing spiking dependence is loglinear modeling
and associated methodology.

Three complications make the problem challenging, at least in
principle. First, there is nonstationarity: the probabilities vary
across time. The data thus form a~sequence of $2^\nu$ contingency
tables. Second, absent from the above notation is a possible dependence
on spiking history. Such dependence is the rule rather than the
exception.  Let $\bar{\mathcal H}^i_t$ denote the set of values of $X^i
(s)$, where $s<t$, and $s, t$ are multiples of $\delta$.  Thus,
$\bar{\mathcal H}_t = (\bar{\mathcal H}^1_t, \ldots, \bar{\mathcal H}^\nu_t)$ is
the history of the binned spike train up to time $t$.  We may wish to
consider conditional probabilities such as
\[
P_{a, b, c}^{1, 2, 3} ( t| \bar{\mathcal H}_t) = P \bigl( X^1 (t) = a, X^2
(t)=b, X^3 (t) =c| \bar{\mathcal H}_t \bigr)
\]
for $a,b,c \in \{0,1\}$. Third, there is the possibility of precisely
timed lagged dependence (or time-delayed synchrony): for example, we
may want to consider the probability
\begin{equation}\label{eq:1.35}
P_{1, 1, 1}^{1, 2, 3} ( s, t, u) = P \bigl( X^1 ( s) = 1, X^2 (t) = 1,
X^3 ( u) = 1 \bigr),
\end{equation}
where $s, t, u$ may be distinct. Similarly, we might consider the
conditional probability
\[
P^{1, 2, 3}_{1, 1, 1} ( s, t, u| \bar{\mathcal H}_s^1, \bar{\mathcal H}_t^2,
\bar{\mathcal H}_u^3 ) = P \bigl( X^1( s) = 1, X^2 ( t) = 1, X^3( u) = 1 |
\bar{\mathcal H}_s^1, \bar{\mathcal H}_t^2, \bar{\mathcal H}_u^3 \bigr).
\]
In principle, we would want to consider all possible combinations of
lags.  Even for $\nu=3$ neurons, but especially as we contemplate
$\nu\gg 3$, strong restrictions must be imposed in order to have any
hope of estimating all these probabilities from relatively sparse data
in a small number of repeated trials. To reduce model dimensionality,
we suggest four seemingly reasonable tactics: (i) considering models
with only low-order interactions, (ii) assuming the probabilities\vspace*{-2pt}
$P^{1, 2, 3}_{a, b, c} (t)$ or $P_{a, b, c}^{1, 2, 3} (t| \bar{\mathcal H}_t)$
vary smoothly across time $t$, (iii)~restricting history effects to
those that modify a neuron's spiking behavior based on its own past
spiking, and then (iv) applying analogues to standard loglinear model
methodology. Combining these, we obtain tractable models for multiple
binary time series to which standard methodology, such as maximum
likelihood and smoothing, may be applied. In modeling synchronous
spiking events as loglinear time series, however, it would be highly
desirable to have a continuous-time representation, where binning
becomes an acknowledged approximation.  We therefore also provide a
theoretical point process foundation for the discrete multivariate
methods proposed here.

It is important to distinguish the probabilities $P_{a, b, c}^{1, 2,
  3} ( t)$ and $P_{a, b, c}^{1, 2, 3} ( t| \bar{\mathcal
  H}_t)$.  The former are trial-averaged or \textit{marginal} probabilities, while the latter are within-trial or
  \textit{conditional} probabilities. Both might be of interest but they
quantify different things.  As an extreme example suppose, as sometimes
is observed, each of two neurons has highly rhythmic spiking at an
approximately constant phase relationship with an oscillatory potential
produced by some large network of cells.  Marginally these neurons will
show strongly dependent spiking. On the other hand, after taking
account of the oscillatory rhythm by conditioning on each neuron's
spiking history and/or a~suitable within-trial time-varying covariate,
that dependence may vanish. Such a~finding would be informative, as it
would clearly indicate the nature of the dependence between the
neurons. In Section \ref{sec:analysis} we give a less dramatic but
similar example taken from the data described in
Section \ref{sec:example}.

We treat marginal and conditional analyses separately. Our use of two
distinct frameworks is a consequence of the way time resolution will
affect continuous-time approximations. We might begin by imagining the
situation in which event times could be determined with infinite
precision. In this case it is natural to assume, as is common in the
point process literature, that no two processes have simultaneous
events. As we indicate, this conception may be applied to marginal
analysis.  However, the event times are necessarily recorded to fixed
accuracy, which becomes the minimal value of $\delta$, and $\delta$ may
be sufficiently large that simultaneous events become a practical
possibility. Many recording devices, for example, store neural spike
event times with an accuracy of 1 millisecond. Furthermore, the time
scale of physiological synchrony---the proximity of spike events
thought to be physiologically meaningful---is often considered to be on
the order of $\delta=5$ milliseconds [cf.  Gr\"un, Diesmann and Aertsen
(\citeyear{GDA2002a}, \citeyear{GDA2002b}) and Gr\"un (\citeyear{Grun2009})]. For within-trial analyses of
synchrony, the theoretical conception of simultaneous (or synchronous)
spikes across multiple trials therefore becomes important and leads us
to the formalism detailed below. The framework we consider here
provides one way of capturing the notion that events within $\delta$
milliseconds of each other are essentially synchronous.

\begin{figure}

\includegraphics{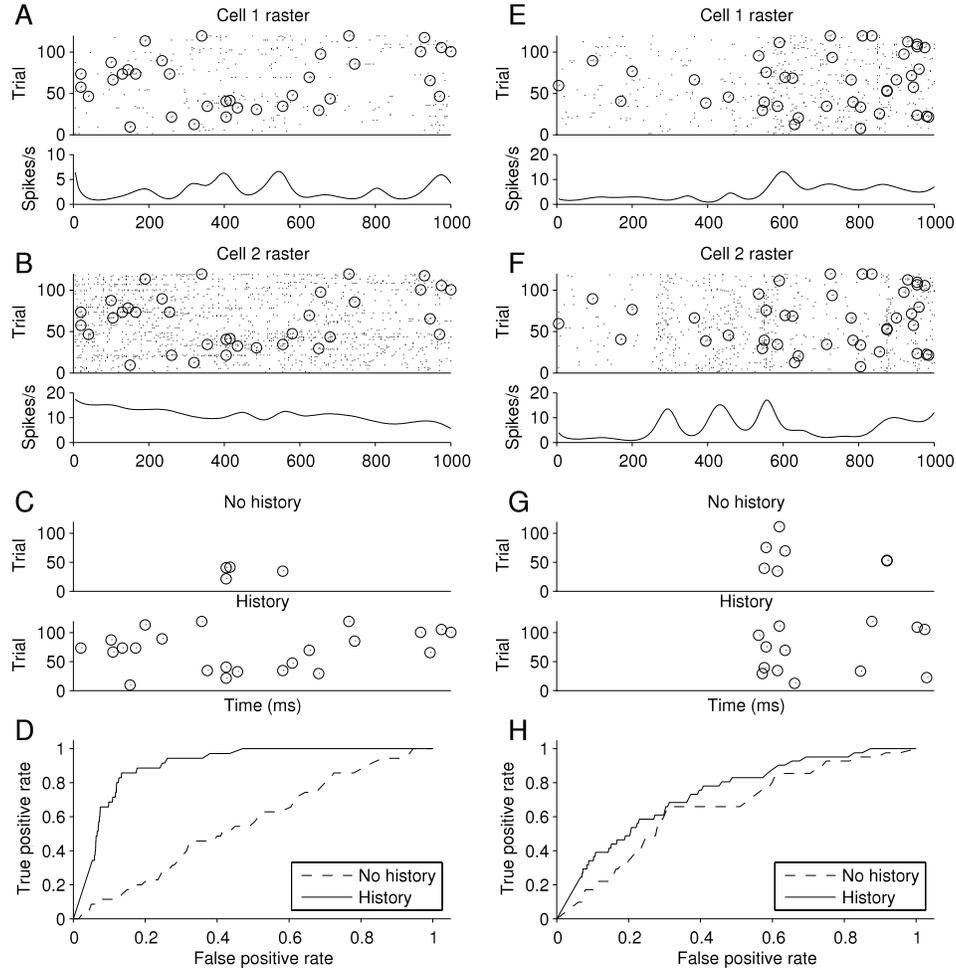}

 \caption{Synchronous spike analysis for two pairs of neurons. Results for one
pair shown on left, in parts \textup{(A)--(D)} and for the other pair on the
right in parts \textup{(E)--(H)}. Part \textup{(A)} Response of a cell to repetitions of
a 1 second drifting grating stimulus.
 The raster plot is shown above and the smoothed PSTH below. Part \textup{(B)} Response
from a second cell, as in \textup{(A)}. In both \textup{(A)} and \textup{(B)}, spikes that are
synchronous between the pair are circled. Part \textup{(C)} Correct joint spike
predictions from model, shown as circles [as in parts \textup{(A)} and \textup{(B)}],
when false positive rate is set at 10\%. In top plot the joint spikes
are from the history-independent model, as in (\protect\ref{loglin2.marginal}),
while in the bottom plot they are as in (\protect\ref{loglin2.conditional}),
including the network covariate in the history term. Part \textup{(D)} ROC
curves for the models in part~\textup{(C)}. Parts \textup{(E)}, \textup{(F)}, \textup{(G)} and \textup{(H)} are
similar to Parts \textup{(A), (B), (C)} and \textup{(D)} but for the second pair of
neurons.} \label{fig2}
\end{figure}

The rest of this article is organized as follows. Section \ref{sec2} presents
the methodology in three subsections: Sections \ref{sec21} and \ref{sec22} introduce
marginal and conditional methods in the simplest case, while
Section \ref{sec:loglinear} discusses the use of loglinear models and
associated methodology for analyzing spiking dependence. In
Section \ref{sec:analysis} we illustrate the methodology by returning
to the example of Section \ref{sec:example}. The main purpose of our
approach is to allow covariates to take account of such things as the
irregular network rhythm displayed in Figure \ref{fig1}, so that
synchrony can be understood as either related to the network effects or
unrelated. Figure \ref{fig2} displays synchronous spiking events for
two different pairs of neurons, together with accompanying fits from
continuous-time loglinear models. For both pairs the independence model
fails to account for synchronous spiking. However, for one pair the
apparent excess synchrony disappears when  history effects are included
in the loglinear model, while in the other pair they do not, leading to
the conclusion that in the second case the excess synchrony must have
some other source. Theory is presented in  Sections \ref{sec:process}--\ref{sec:conditional}. We add some
discussion in Section \ref{sec7}. All proofs in this article are deferred to the
\hyperref[appendix]{Appendix}.

\section{Methodology}\label{sec2}

In this section we present our continuous-time loglinear modeling
methodology. We begin with the simplest case of $\nu=2$ neurons,
presenting the main ideas in Sections \ref{sec21} and \ref{sec22} for the marginal and
conditional cases, respectively, in terms of the probabilities $P^1_a(
t) = P( X^1( t)= a)$, $P^{1,2}_{a,b} (t) =P(X^1 (t) =a,X^2 (t) =b)$,
etc.,  for all $a,b \in \{0, 1\}$. We show how we wish to pass to the
continuous-time limit, thereby introducing point process technology and
making sense of continuous-time smoothing, which is an essential
feature of our approach. In Section \ref{sec:loglinear} we reformulate using loglinear
models, and then give continuous-time loglinear models for $\nu=3$. Our
analyses in Section \ref{sec:analysis} are confined to $\nu=2$ and
$\nu=3$ because of the paucity of higher-order synchronous spikes in
our data. Our explicit models for $\nu=3$ should make clear how
higher-order models are created. We give general recursive formulas in
Sections \ref{sec:marginal} and \ref{sec:conditional}.

\subsection{Marginal methods for $\nu=2$}\label{sec21}

The null hypothesis
\begin{equation}\label{eq:2.1}
H_0 \dvtx P^{1,2}_{1,1} (t) = P_1^1 (t) P^2_1 (t)\qquad\forall
t\in {\mathcal T} = \{0, \delta, 2\delta, \ldots, T-\delta\}
\end{equation}
is a statement that both neurons spike in the interval $[t, t+\delta)$,
on the average, at the rate determined by independence. Defining $\zeta
(t)$ by
%
\begin{equation}\label{eq:2.7}
  P^{1, 2}_{1, 1} (t) =
P^1_1 (t) P^2_1(t) \zeta (t) \qquad \forall t\in {\mathcal T},
\end{equation}
we may rewrite (\ref{eq:2.1}) as
%
\begin{equation}
H_0\dvtx \zeta (t) =1 \qquad\forall t\in {\mathcal T}. \label{eq:2.63}
\end{equation}
As in Ventura, Cai and Kass (\citeyear{VCK2005}),
 to assess $H_0$, the general strategy we follow is to (i) smooth
the observed-frequency estimates of $P^1_1 (t)$, $P^2_1 (t)$ and $P^{1,
2}_{1, 1} (t)$ across time $t$, and then (ii) form a suitable test
statistic and compute a $p$-value using a bootstrap procedure. We may
deal with time-lagged hypotheses similarly, for example, for a lag
$h>0$, we write
\begin{eqnarray}\label{eq:2.42}
P^{1, 2}_{1, 1} (t, t + \delta h) &=& P \bigl( X^1 (t) = 1, X^2(t + \delta
h) =1 \bigr)\nonumber\\ [-8pt]\\ [-8pt]
 &=& P^1_1 (t) P^2_1 (t+ \delta h) \zeta(t, t +\delta h),\nonumber
\end{eqnarray}
then smooth the observed-frequency estimates for $P^{1, 2}_{1, 1} (t,
t+\delta h)$ as a function of~$t$, form an analogous test statistic and
find a $p$-value.

To formalize this approach, we consider counting processes $N^i_t$
corresponding to the point processes $s_1^i, s_2^i,\ldots,
s^i_{N^i_t}$, $i= 1,2$ (as in Section \ref{sec:overview} with $\nu=2$).  Under
regularity conditions, the following limits exist:
\begin{eqnarray}\label{eq:2.2}
\lambda^i (t) &=& \lim_{\delta \rightarrow 0} \delta^{-1} P\bigl( N^i_{t
+\delta}  - N^i_t  = 1\bigr),\nonumber \\
\lambda^{1, 2} (t) &=& \lim_{\delta \rightarrow 0} \delta^{-2} P \bigl(
(N^1_{t + \delta} - N^1_t ) (N^2_{t + \delta} - N^2_t ) =1 \bigr),\\
\xi (t) &=& \lim_{\delta \rightarrow 0} \zeta (t). \nonumber
\end{eqnarray}
Consequently, for small $\delta$, we have
\[
P^i_1 (t) \approx \lambda^i(t) \delta, \qquad P^{1, 2}_{1, 1}
(t) \approx \lambda^{1, 2} (t) \delta^2.
\]
The smoothing of the observed-frequency estimates for $P_{1, 1}^{1, 2}
(t)$ may be understood as producing an estimate $\hat{\lambda}^{1, 2}
(t)$ for $\lambda^{1, 2}(t)$. The null hypothesis in~(\ref{eq:2.1})
becomes, in the limit as $\delta\rightarrow 0$,
\[
H_0\dvtx \lambda^{1, 2}(t) = \lambda^1 (t) \lambda^2(t)\qquad
\forall t\in [0, T),
\]
or, equivalently,
\begin{equation}\label{H0xi1}
H_0: \xi (t) = 1\qquad \forall t\in [0, T).
\end{equation}
The lag $h$ case is treated similarly. Under mild  conditions, Theorems
\ref{tm:4.3} and \ref{tm:4.2} of Section \ref{sec:marginal} show that the above heuristic
arguments hold for  a continuous-time regular marked point process.
This in turn  gives a rigorous  asymptotic justification (as $\delta
\rightarrow 0$) for estimation and testing procedures such as those in
steps (i) and (ii) mentioned above, following (\ref{eq:2.63}), and
illustrated in Section~\ref{sec:analysis}.

\subsection{Conditional methods for $\nu=2$}\label{sec22}

To deal with history effects, equation (\ref{eq:2.7}) is replaced with
\begin{equation}\label{eq:2.8}
  P^{1, 2}_{1, 1} (t| \bar{\mathcal H}_t) =
P^1_1 (t | \bar{\mathcal H}^1_t) P^2_1 (t | \bar{\mathcal H}^2_t)\zeta (t)
\qquad \forall t\in {\mathcal T},
\end{equation}
where $\bar{\mathcal H}^i_t$, $i=1,2$, are, as in Section \ref{sec1}, the binned
spiking histories of neurons 1 and 2, respectively, on the interval
$[0, t)$. Analogous to (\ref{eq:2.63}), the null hypothesis is
\[
H_0\dvtx \zeta (t) =1\qquad \forall t\in {\mathcal T}.
\]
We note that there are two substantial simplifications in
(\ref{eq:2.8}). First, $P^i_1(t| \bar{\mathcal H}_t) = P^i_1 (t | \bar{\mathcal
H}^i_t)$, which says that neuron $i$'s own history $\bar{\mathcal H}^i_t$
is relevant in modifying its spiking probability (but not the other
neuron's history).  Second, $\zeta(t)$ does not depend on the spiking
history $\bar{\mathcal H}_t$.  This is important for what it claims about
the physiology, for the way it simplifies statistical analysis, and for
the constraint it places on the point process framework.
Physiologically, it decomposes excess spiking into history-related
effects and stimulus-related effects, which allows the kind of
interpretation alluded to in Section \ref{sec1} and presented in our data
analysis in Section~\ref{sec:analysis}. Statistically, it improves
power because tests of $H_0$ effectively pool information across
trials, thereby increasing the effective sample size.

Consider counting processes $N^i_t$, $i=1,2$, as in Section \ref{sec21}. Under
regularity conditions, the following limits exist for $t\in [0, T)$:
\begin{eqnarray} \label{eq:2.83}
\lambda^i ( t| {\mathcal H}^i_t) &=& \lim_{\delta \rightarrow 0}
\delta^{-1} P( N^i_{t+\delta} - N^i_t =1| \bar{\mathcal H}_t^i),\qquad
 i=1,2,\nonumber \\
\lambda^{1, 2} ( t| {\mathcal H}_t) &=& \lim_{\delta \rightarrow 0}
\delta^{-2} P\bigl( (N^1_{t +\delta} - N^1_t)  (N^2_{t+\delta} - N^2_t)
=1| \bar{\mathcal H}_t \bigr),\\
\xi ( t) &=& \lim_{\delta \rightarrow 0} \zeta (t), \nonumber
\end{eqnarray}
where ${\mathcal H}_t = \lim_{\delta\rightarrow 0} \bar{\mathcal H}_t$ and
${\mathcal H}_t^i = \lim_{\delta\rightarrow 0} \bar{\mathcal H}_t^i, i=1,2$.
For sufficiently small $\delta$, we have
%
\begin{equation} \label{eq:sparse2}
\hspace*{25pt}P^i_1 (t | \bar{\mathcal H}^i_t) \approx \lambda^i( t| {\mathcal H}^i_t)
\delta,\qquad i=1,2,\quad \mbox{and}\quad P^{1, 2}_{1, 1} (t | \bar{\mathcal H}_t) \approx
\lambda^{1, 2}( t| {\mathcal H}_t) \delta^2
\end{equation}
for all $t\in {\mathcal T}$. Again following Ventura, Cai and Kass (\citeyear{VCK2005}), we may
smooth the observed-frequency estimates of $P^{1, 2}_{1, 1} (t|
\bar{\mathcal H}_t)$ to produce an estimate of $\lambda^{1, 2} (t| {\mathcal
H}_t)$, and smooth the observed-frequency estimates of $P^i_1(t|
\bar{\mathcal
  H}^i_t)$ to produce estimates of $\lambda^i (t| {\mathcal H}_t^i)$,
$i=1,2$.  Letting $\delta\rightarrow 0$ in (\ref{eq:2.8}), we obtain
\begin{equation}\label{eq:2.84}
\lambda^{1, 2} (t | {\mathcal H}_t) = \xi (t) \lambda^1 (t|{\mathcal H}_t^1)
\lambda^2 (t| {\mathcal H}_t^2) \qquad\forall t\in [0, T).
\end{equation}
Consequently, for sufficiently small $\delta$, a conditional test of
$H_0\dvtx \zeta (t) =1$ for all~$ t$ becomes a test of the null hypothesis
$H_0 \dvtx \lambda^{1, 2} (t | {\mathcal H}_t) = \lambda^1 (t|{\mathcal H}_t^1)
\lambda^2 (t| {\mathcal H}_t^2)$ for all $t$ or, equivalently, in this
conditional case we have the same null hypothetical statement as
(\ref{H0xi1}).

In attempting to make equation (\ref{eq:2.83}) rigorous, a difficulty
arises: for a~regular marked point process, the function $\xi$ need not
be independent of the spiking history. This would create a fundamental
mismatch between the discrete data-analytical method and its
continuous-time limit. The key to avoiding this problem is to enforce
the sparsity condition (\ref{eq:sparse2}). Specifically, the
probabilities $P^i_1 (t | \bar{\mathcal H}^i_t)$ are of  order $O(\delta)$,
while the probabilities $P^{1, 2}_{1, 1} (t | \bar{\mathcal H}_t)$ are of
order $O(\delta^2)$. This also allows independence models within the
marked point process framework. Section \ref{sec:conditional} proposes
a class of marked point process models indexed by $\delta$ and provides
results that validate the heuristics above.

\subsection{Loglinear models}\label{sec:loglinear}

We now reformulate in terms of loglinear models the procedures sketched
in Sections \ref{sec21} and \ref{sec22} for $\nu=2$ neurons, and then indicate the way
generalizations proceed when $\nu \ge 3$.

In the marginal case of Section \ref{sec21}, it is convenient to define
\begin{eqnarray*}
\tilde{P}^{1,2}_{0,0} (t) &=& 1,\\
\tilde{P}^{1,2}_{1,0} (t) &=& P^1_1 (t),\\
\tilde{P}^{1,2}_{0,1} (t) &=& P^2_1 (t),\\
\tilde{P}^{1,2}_{1,1} (t) &=& P^{1,2}_{1,1} (t)\qquad\forall
t\in {\mathcal T}.
\end{eqnarray*}
Equation (\ref{eq:2.7}) implies that
%
\begin{equation}\label{loglinear2}
\log[ \tilde{P}^{1, 2}_{a, b} (t)] =  a \log [ P^1_1 (t) ] +  b \log [
P^2_1 (t) ] + ab  \log [ \zeta( t) ]
\end{equation}
for all $a,b\in \{0,1\}$ and  $t\in {\mathcal T}$ and in the
continuous-time limit, using (\ref{eq:2.2}), we write
%
\begin{equation} \label{loglin2.marginal}
\log \lambda^{1, 2} ( t) = \log \lambda^1 ( t) + \log \lambda^2 ( t) +
\log [ \xi( t) ]
\end{equation}
for $t\in [0, T)$. The null hypothesis may then be written as
%
\begin{equation}\label{eq:H0log.xi}
H_0\dvtx \log [ \xi( t) ] =0\qquad\forall  t\in [0, T).
\end{equation}
In the conditional case of Section 2.2, we similarly define
\begin{eqnarray*}
\tilde{P}^{1,2}_{0,0} (t| \bar{\mathcal H}_t ) &=& 1,\\
\tilde{P}^{1,2}_{1,0} (t| \bar{\mathcal H}_t   ) &=& P^1_1 (t | \bar{\mathcal
H}_t),\\
\tilde{P}^{1,2}_{0,1} (t| \bar{\mathcal H}_t) &=& P^2_1 (t| \bar{\mathcal
H}_t), \\
\tilde{P}^{1,2}_{1,1} (t| \bar{\mathcal H}_t) &=& P^{1,2}_{1,1} (t|
\bar{\mathcal H}_t)\qquad\forall t\in {\mathcal T},
\end{eqnarray*}
we may rewrite (\ref{eq:2.8}) as the loglinear model
%
\begin{equation} \label{loglinear2H}
\log[ \tilde{P}^{1, 2}_{a, b} (t | \bar{\mathcal H}_t)] =  a \log [ P^1_1
(t| \bar{\mathcal H}_t^1) ] +  b \log [ P^2_1 (t| \bar{\mathcal H}_t^2) ] + ab
\log [ \zeta( t) ]
\end{equation}
for all $a,b\in \{0,1\}$ and  $t\in {\mathcal T}$ and in the
continuous-time limit we rewrite (\ref{eq:2.84}) in the form
%
\begin{equation}\label{loglin2.conditional}
\log \lambda^{1, 2} ( t| {\mathcal H}_t) = \log \lambda^1 ( t| {\mathcal
H}^1_t) + \log \lambda^2 ( t| {\mathcal H}^2_t) + \log [ \xi( t) ]
\end{equation}
for $t\in [0, T)$. The null hypothesis may again be written as in
(\ref{eq:H0log.xi}).

Rewriting the model in loglinear forms (\ref{loglinear2}),
(\ref{loglin2.marginal}), (\ref{loglinear2H}) and
(\ref{loglin2.conditional}) allows us to generalize to $\nu\geq 3$
neurons. For example, with the obvious extensions of the previous
definitions, for $\nu=3$ neurons the two-way interaction model in the
continuous-time marginal case becomes
%
\begin{eqnarray}\label{loglinear3}
\hspace*{25pt}\log[ \tilde P^{1, 2,3}_{a,b,c} (t)] &=&   \log [ P^1_1 (t) ] +   \log
[ P^2_1 (t) ] +    \log [ P^3_1 (t) ]\nonumber\\ [-8pt]\\ [-8pt]
&&{} + ab  \log \bigl[ \zeta_{\{1,2\}}( t) \bigr] + ac  \log \bigl[ \zeta_{\{1,3\}}( t)
\bigr] + bc  \log \bigl[ \zeta_{\{2,3\}}( t) \bigr]\nonumber
\end{eqnarray}
for all $a,b,c \in \{0,1\}$ and  $t\in {\mathcal T}$, and
\begin{eqnarray*}
\log[ \lambda^{1, 2,3} (t)] &=&   \log [ \lambda^1 (t) ] +   \log [
\lambda^2 (t) ] +    \log [ \lambda^3 (t) ]\\
&&{} +   \log \bigl[ \xi_{\{1,2\}}( t) \bigr] +   \log \bigl[ \xi_{\{1,3\}}( t) \bigr] +
\log \bigl[ \xi_{\{2,3\}}( t) \bigr]\nonumber
\end{eqnarray*}
for all $t\in (0,T]$. The general form of (\ref{loglinear3}) is given
by equation (\ref{eq:4.91}) in Section~\ref{sec:marginal}. In the
conditional case, the two-way interaction model becomes
\begin{eqnarray}\label{loglinear3H}
\hspace*{23pt}\log[ \tilde{P}^{1, 2,3}_{a, b,c} (t|\bar{\mathcal H}_t)] &=&  a \log [
P^1_1 (t|\bar{\mathcal H}_t) ] +  b \log [ P^2_1 (t|\bar{\mathcal H}_t) ] + c
\log [ P^3_1 (t|\bar{\mathcal H}_t) ]\nonumber\\ [-8pt]\\ [-8pt]
&&{} + ab  \log \bigl[ \zeta_{\{1,2\}}( t) \bigr] + ac  \log \bigl[ \zeta_{\{1,3\}}( t)
\bigr] + bc  \log \bigl[ \zeta_{\{2,3\}}( t) \bigr]\nonumber
\end{eqnarray}
for all $a,b,c \in \{0,1\}$ and  $t\in {\mathcal T}$ and in continuous
time,
\begin{eqnarray*}
\log[ \lambda^{1, 2,3} (t|{\mathcal H}_t)] &=&   \log [ \lambda^1 (t|{\mathcal
H}_t) ] +   \log [ \lambda^2 (t|{\mathcal H}_t) ] +    \log [ \lambda^3
(t|{\mathcal H}_t) ]\\
&&{} +   \log \bigl[ \xi_{\{1,2\}}( t) \bigr] +   \log\bigl [ \xi_{\{1,3\}}( t) \bigr] +
\log \bigl[ \xi_{\{2,3\}}( t) \bigr]
\end{eqnarray*}
for all $t\in (0,T]$. In either the marginal or conditional case, the
null hypothesis of independence may be written as
\begin{equation}\label{H0indep3}
H_0\dvtx \log \bigl[ \xi_{\{i,j\}}( t) \bigr] =0\qquad \forall t\in (0, T],
1\leq i < j \leq 3.
\end{equation}
On the other hand, we could include the additional term $abc \log [
\xi_{\{1,2,3\}}( t) ]$ and use the null hypothesis of no three-way
interaction
\begin{equation}\label{H02way}
H_0\dvtx \log \bigl[ \xi_{\{1,2,3\}}( t) \bigr]=0 \qquad \forall t\in (0, T].
\end{equation}

These loglinear models offer a simple and powerful way to study
dependence among neurons when spiking history is taken into account.
They have an important dimensionality reduction property in that the
higher-order terms are asymptotically independent of history.  In
practice, this provides a huge advantage: the synchronous spikes are
relatively rare; in assessing excess synchronous spiking with this
model, the data may be pooled over different histories, leading to a
much larger effective sample size. The general conditional model in
equation (\ref{eq:5.76}) retains this structure. An additional feature
of these loglinear models is that time-varying covariates may be
included without introducing new complications. In
Section \ref{sec:analysis} we use a covariate to characterize the
network up states, which are visible in part~C of Figure \ref{fig1},
simply by including it in calculating each of the individual-neuron
conditional\vadjust{\goodbreak} intensities $\lambda^1(t|{\mathcal H}^1_t)$ and
$\lambda^2(t|{\mathcal H}^2_t)$ in (\ref{loglin2.conditional}).

Sometimes, as in the data we analyze here, the synchronous events are
too sparse to allow estimation of time-varying excess synchrony and we
must assume it to be constant, $\zeta(t)=\zeta$ for all $t$.  Thus, for
$\nu=2$, the models of (\ref{loglinear2}) or (\ref{loglinear2H}) take
simplified forms in which $\zeta(t)$ is replaced by the constant
$\zeta$ and we would use different test statistics to test the null
hypothesis $H_0\dvtx \zeta=1$. To distinguish the marginal and conditional
cases, we replace $\zeta(t)$ by $\zeta_H$ in (\ref{loglinear2H}) and
then also write $H_0\dvtx \zeta_H=1$. Moving to continuous time, which is
simpler computationally, we write $\xi(t)=\xi$, estimate $\xi$ and
$\xi_H$, and test $H_0\dvtx\xi=1$  and $H_0\dvtx\xi_H=1$.  Specifically, we
apply the loglinear models (\ref{loglinear2}),
(\ref{loglin2.marginal}), (\ref{loglinear2H}) and
(\ref{loglin2.conditional}) in two steps. First, we smooth the
respective PSTHs to produce smoothed curves $\hat \lambda^i(t)$, as in
parts A and B of Figure \ref{fig1}. Second, ignoring estimation
uncertainty and taking $\lambda^i(t)=\hat \lambda^i(t)$, we estimate
the constant $\zeta$. Using the point process representation of joint
spiking (justified by the results in Sections \ref{sec:marginal} and
\ref{sec:conditional}), we may then write
\[
\log L(\xi) = -\int \lambda(t)\,dt + \sum \log \lambda(t_i),
\]
where the sum is over the joint spike times $t_i$ and $\lambda(t)$ is
replaced by the right-hand side of (\ref{loglin2.marginal}), in the
marginal case, or (\ref{loglin2.conditional}), in the conditional case.
It is easy to maximize the likelihood $L(\xi)$ analytically: setting
the left-hand side to $\ell(\xi)$, in the marginal case we have
\[
\ell^{\prime}(\xi) = - \int \lambda^1(t)\lambda^2(t)\,dt +\frac{N}{\xi},
\]
where $N$ is the number of joint (synchronous) spikes (the number of
terms in the sum), while in the conditional case we have the analogous
formula
\[
\ell^{\prime}(\xi_H) = - \int \lambda^1(t|{\mathcal
H}_t^1)\lambda^2(t|{\mathcal H}_t^2)\,dt +\frac{N}{\xi_H}
\]
and setting to 0 and solving gives
\begin{equation}\label{eq:zetahat}
\hat \xi = \frac{N}{\int \lambda^1(t)\lambda^2(t)\,dt}
\end{equation}
and
\begin{equation} \label{eq:zetahatH}
\hat \xi_H = \frac{N}{\int \lambda^1(t|{\mathcal H}_t^1)\lambda^2(t|{\mathcal
H}_t^2)\,dt},
\end{equation}
which, in both cases, is the ratio of the number of observed joint
spikes to the number expected under independence.

We apply (\ref{eq:zetahat}) and (\ref{eq:zetahatH}) in
Section \ref{sec:analysis}. To test $H_0\dvtx \xi=1$ and $H_0\dvtx \xi_H=1$, we
use a bootstrap procedure in which we generate spike trains under the
relevant null-hypothetical model. This is carried out in discrete time,
and requires all 4 cell probabilities $\tilde P^{1,2}_{a,b}(t)$ or
$\tilde P^{1,2}_{a,b}(t|\bar{\mathcal H}_t)$ at every time $t \in {\mathcal
T}$. These are easily obtained by subtraction using
$P^1_1(t)=\lambda^1(t)\delta$, $P^2_1(t)=\lambda^2(t)\delta$, and $\hat
\zeta=\hat \xi$ or, in the conditional case, $P^1_1(t|\bar{\mathcal
H}_t^1)=\lambda^1(t|{\mathcal H}^1_t)\delta$, $P^2_1(t|\bar{\mathcal
H}_t^2)=\lambda^2(t|{\mathcal H}^2_t)\delta$, and $\hat \zeta_H = \hat
\xi_H$. As we said above, $\lambda^i(t)=\hat \lambda^i(t)$ is
 obtained from the preliminary step of smoothing the PSTH.
Similarly, the conditional intensities $\lambda^i(t|{\mathcal H}^i_t)=\hat
\lambda^i(t|{\mathcal H}^i_t)$ are obtained from smooth history-dependent
intensity models such as those discussed in Kass, Ventura and Brown (\citeyear{KVB2005}). In the
analyses reported here we have used fixed-knot splines to describe
variation across time $t$.

In the case of 3 or more neurons the analogous estimates and cell
probabilities must, in general, be obtained by a version of iterative
proportional fitting. For $\nu=3$, to test the null hypothesis
(\ref{H02way}), we follow the steps leading to (\ref{eq:zetahat}) and
(\ref{eq:zetahatH}). Under the assumption of constant $\zeta_{123}$, we
obtain
\begin{equation}\label{eq:zetahat123}
\hat \xi_{123} = \frac{N}{\int
  \lambda^1(t)\lambda^2(t)\lambda^3(t)
\xi_{12}(t)\xi_{13}(t)\xi_{23}(t)\,dt }
\end{equation}
and
\begin{equation}\label{eq:zetahat123H}
\hat \xi_{123,H} = \frac{N}{\int \lambda^1(t|{\mathcal
H}_t^1)\lambda^2(t|{\mathcal H}_t^2) \lambda^3(t|{\mathcal H}_t^3)
\xi_{12,H}(t)\xi_{13,H}(t)\xi_{23,H}(t)\,dt}.
\end{equation}
In Section \ref{sec:analysis} we fit (\ref{loglinear3}) and report a
bootstrap test of the hypothesis (\ref{H02way}) using the test
statistic $\hat \xi_{123}$ in (\ref{eq:zetahat123}).

\section{Data analysis}\label{sec:analysis}

We applied the methods of Section \ref{sec:loglinear} to a subset of
the data described in Section \ref{sec:example} and present the results
here. We plan to report a more comprehensive analysis elsewhere.

We took $\delta = 5$ milliseconds (ms), which is a commonly-used window
width in studies of synchronous spiking. Raster plots of spike trains
across repeated trials from a pair of neurons are shown in Parts A and
B of Figure \ref{fig2}, with synchronous events indicated by circles.
Below each raster plot is a smoothed PSTH, that is, the two plots show
smoothed estimates $\hat \lambda^1(t)$ and $\hat \lambda^2(t)$ of
$\lambda^1(t)$ and $\lambda^2(t)$ in (\ref{eq:2.2}), the units being
spikes per second. Smoothing was performed by fitting a generalized
additive model using cubic splines with knots spaced 100 ms apart.
Specifically, we applied Poisson regression to the count data resulting
from pooling the binary spike indicators across trials: for each time
bin the count was the number of trials on which a spike occurred. To
test $H_0$ under the model in (\ref{loglin2.marginal}), we applied
(\ref{eq:zetahat}) to find $\log \hat \xi$. We then computed a
parametric bootstrap standard error of $\log \hat \xi$ by generating
pseudo-data from model (\ref{loglin2.marginal}) assuming $H_0\dvtx \log
\xi=0$. We generated 1000 such trials, giving 1000 pseudo-data values
of $\log \hat \xi$, and computed the standard deviation of those values
as a standard error, to obtain an observed $z$-ratio test statistic of
3.03 ($p=0.0012$ according to asymptotic normality).

The highly significant $z$ ratio shows that there is excess sychronous
spiking beyond what would be expected from the varying firing rates of
the two neurons under independence. However, it does not address the
source of the excess synchronous spiking. The excess synchronous
spiking could depend on the stimulus or, alternatively, it might be due
to the slow waves of population activity evident in part (C) of Figure
1, the time of which vary from trial to trial and therefore do not
depend on the stimulus. To examine the latter possibility, we applied a
within-trial loglinear model as in (\ref{loglin2.conditional}) except
that we incorporated into the history effect not only the history of
each neuron but also a covariate representing the population effect.
Specifically, for neuron $i$ ($i=1,2$) we used the same generalized
additive model as before, but with two additional variables. The first
was a variable that, for each time bin, was equal to the number of
neuron $i$ spikes that had occurred in the previous 100 ms. The second
was a variable that, for each time bin, was equal to the number of
spikes that occurred in the previous 100 ms across the whole population
of neurons, other than neurons 1 and 2. We thereby obtained fitted
estimates $\hat \lambda^1 ( t| {\mathcal H}^1_t)$ and $\hat \lambda^2 ( t|
{\mathcal H}^2_t)$ of $\lambda^1 ( t| {\mathcal H}^1_t)$ and $\lambda^2 ( t|
{\mathcal H}^2_t)$. Note that the fits for the independence model, defined
by applying (\ref{H0xi1}) to (\ref{eq:2.84}), now vary from trial to
trial due to the history effects. Applying (\ref{eq:zetahatH}), we
found $\log \hat \xi_H $, and then again computed a bootstrap standard
error of $\log \hat \xi_H$ by creating 1000 trials of pseudo-data, giving
$\log \hat \xi_H=0.06\pm 0.15$, for a~$z$-ratio of $0.39$, which is clearly not significant.

Raster plots for a different pair of neurons are shown in parts (E) and
(F) of Figure \ref{fig2}. The same procedures were applied to this
pair. Here, the $z$-ratio for testing $H_0$ under the marginal model
was 3.77 ($p<0.0001$), while that for testing $H_0$ under the
conditional  model remained highly significant at 3.57 ($p=0.0002$) with $\log \hat \xi_H=0.82\pm 0.23$. In
other words, using the loglinear model methodology, we have discovered
two pairs of V1 neurons with quite different behavior. For the first
pair, synchrony can be explained entirely by network effects, while for
the second pair it can not; this suggests that, instead, for the second
pair, some of the excess synchrony may be stimulus-related.

We also compared the marginal and conditional models
(\ref{loglin2.marginal}) and (\ref{loglin2.conditional}) using ROC
curves. Specifically, for the binary joint spiking data we used each
model to predict a~spike whenever the intensity was larger than a~given
constant: for the marginal case whenever $\log \lambda^{1, 2} ( t) >
c_{\mathrm{marginal}}$, and for the conditional case whenever $\log \lambda^{1,
2} ( t| {\mathcal H}_t) > c_{\mathrm{conditional}}$. The choice of constants
$c_{\mathrm{marginal}}$ and $c_{\mathrm{conditional}}$ reflect trade-offs between false
positive and true positive rates (analogous to type I error and power)
and as we vary the constants, the plot of true vs. false positive rates
forms the ROC curve. To determine the true and false positive rates, we
performed ten-fold cross-validation, repeatedly fitting from 90\% of
the trials and predicting from the remaining 10\% of the trials. The
two resulting ROC curves are shown in part D of Figure \ref{fig2},
labeled as ``no history'' and ``history,'' respectively. To be clear,
in the two cases we included the terms corresponding, respectively, to
$\xi$ and $\xi_H,$ and in the history case we included both the
auto-history and the network history variables specified above. The ROC
curve for the conditional model strongly dominates that for the
marginal model, indicating far better predictive power. In part~C of
Figure \ref{fig2} we display the true positive joint spike predictions
when the false-positive rate was held at 10\%. These
correctly-predicted joint spikes may be compared to the complete set
displayed in parts A and B of the figure. The top display in part C,
labeled ``no history,'' shows that only a few joint spikes were
correctly predicted by the marginal model, while the large majority
were correctly predicted by the conditional model. Furthermore, the
correctly predicted joint spikes are spread fairly evenly across time.
In contrast, the ROC curves for the second pair of neurons, shown in
part (G) of Figure \ref{fig2}, are close to each other: inclusion of
the history effects (which were statistically significant) did not
greatly improve predictive power. In (G), the correctly predicted
synchronous spikes are clustered in time, with the main cluster
occurring near a peak in the individual-neuron firing-rate functions
shown in the two smoothed PSTHs in parts (E) and (F).

Taking all of the results together, our analysis suggests that the
first pair of neurons produced excess synchronous spikes solely in
conjunction with network effects, which are unrelated to the stimulus,
while for the second pair of neurons some of the excess synchronous
spikes occurred separately from the network activity and were, instead,
stimulus-related.
\begin{figure}

\includegraphics{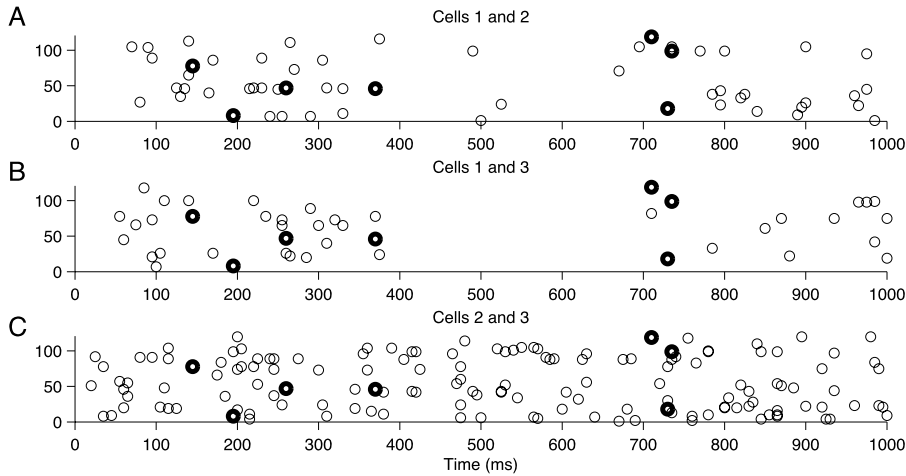}

\caption{Plots of synchronous spiking events for 3 neurons. Each of
the three plots displays all joint spikes (as circles) for a particular
pair of neurons. The dark circles in each plot indicate the 3-way joint
spikes.} \label{fig3}
\end{figure}

We also tried to assess whether 2-way interactions were sufficient to
explain observed 3-way events by fitting the no-3-way interaction model
given by (\ref{loglinear3}), and then testing the null hypothesis in
(\ref{H02way}). We did this  for a~particular set of 3 neurons, whose
joint spikes are displayed in Figure \ref{fig3}.
 The method is analogous to that
carried out above for pairs of neurons, in the sense that the test
statistic was $\hat \xi_{123}$ given by (\ref{eq:zetahat123}) and a
parametric bootstrap procedure, based on the fit of (\ref{loglinear3}),
was used to compute an approximate $p$-value. Fitting of
(\ref{loglinear3}) required an iterative proportional fitting
procedure, which we will describe in detail elsewhere. We obtained
$p=0.16$, indicating no significant 3-way interaction. In other words,
for these three neurons, 2-way excess joint spiking appears able to
explain the occurrence of the 3-way joint spikes. However, as may be
seen in Figure \ref{fig3}, there are very few 3-way spikes in the data.
We mention this issue again in our discussion.


\section{A marked point process framework}\label{sec:process}

In this section a class of marked point processes for modeling neural
spike trains is briefly surveyed. These models take into account the
possibility of two or more neurons firing in synchrony (i.e., at the
same time). Consider an ensemble of $\nu$ neurons labeled~$1$ to $\nu$.
For $T>0$, let $N_T$ denote the total number of spikes produced by this
ensemble on the time interval $[0,T)$ and let $0\leq s_1<\cdots <
s_{N_T} < T$ denote the specific spike times. For each $j=1,\ldots,
N_T$, we write $(s_j, (i_1,\ldots, i_k) )$ to denote the event that a
spike was fired (synchronously) at time $s_j$ by (and only by) the
$i_1$, $\ldots$, $i_k$ neurons. We observe that
\begin{equation}\label{eq:8.1}
{\mathcal H}_T = \{ (s_1, \kappa_1), \ldots, (s_{N_T},
\kappa_{N_T})\dvtx
\kappa_j \in {\mathcal K}, j=1,\ldots, N_T \}
\end{equation}
forms a marked point  process on the interval $[0, T)$ with ${\mathcal K}$
as the mark space satisfying
\[
{\mathcal K} \subseteq \{ (i_1,\ldots, i_k)\dvtx 1\leq i_1<\cdots < i_k\leq
\nu, k=1,\ldots, \nu\}.
\]
We follow Daley and Vere-Jones (\citeyear{DV2002}), page 249, and define the
conditional intensity function of ${\mathcal H}_T$ as
\begin{eqnarray*}
&&\lambda (t,\kappa| {\mathcal H}_t)\\
 &&\qquad= \lambda (t, \kappa| \{ (s_1,
\kappa_1),\ldots, (s_{N_t}, \kappa_{N_t}) \} )
\nonumber \\
&&\qquad= \cases{
h_1(t) f_1(\kappa|t) \qquad \forall 0\leq t\leq s_1, \cr
h_i (t| (s_1, \kappa_1), \ldots, (s_{i-1}, \kappa_{i-1}) ) f_i
(\kappa|t; (s_1, \kappa_1),\ldots, (s_{i-1}, \kappa_{i-1}) )
\cr
\qquad \forall s_{i-1} < t \leq s_i, i=2,\ldots, N_T, \cr
h_{N_T+1} (t| (\!s_1,\kappa_1\!),\ldots, (s_{N_t}, \kappa_{N_T}\!) \!)
f_{N_T+1} (\kappa| t; (s_1,\kappa_1\!),\ldots, (s_{N_T}, \kappa_{N_T}\!) \!)
\cr
\qquad\forall s_{N_T} < t < T,
}
\end{eqnarray*}
where $h_1(\cdot)$ is the hazard function for the location of the first
spike $s_1$, $h_2(\cdot|(s_1,\kappa_1))$ the hazard function for the
location of the second spike $s_2$ conditioned by $(s_1,\kappa_1)$, and
so on, while $f_1(\cdot|t)$ is the conditional probability mass function of
$\kappa_1$ given $s_1=t$, and so on. It is also convenient to write
$\lambda (t,\kappa|\varnothing) = \lambda (t, \kappa| {\mathcal H}_t)$ for
all $t< s_1$. The following proposition and its proof can be found in
Daley and Vere-Jones~(\citeyear{DV2002}), page 251.

\begin{pn} \label{pn:8.1}
Let ${\mathcal H}_T$ be as in  (\ref{eq:8.1}). Then the density of ${\mathcal
H}_T$ is given by
\begin{eqnarray*}
p_\lambda( {\mathcal H}_T)  &=& p_\lambda (\{ (s_1,\kappa_1),\ldots,
(s_{N_T}, \kappa_{N_T}) \})\\
& = &\Biggl[ \prod_{i=1}^{N_T} \lambda (s_i,
\kappa_i | {\mathcal H}_{s_i}) \Biggr] \exp\biggl[ -\sum_{\kappa\in {\mathcal K}} \int_0^T
\lambda (t, \kappa| {\mathcal H}_t)\,dt \biggr].
\end{eqnarray*}
\end{pn}

\section{Theoretical results: Marginal methods}\label{sec:marginal}

In this section we (i) provide a~justification of the limiting
statements in (\ref{eq:2.2}) and (ii) generalize to higher-order
models. We also note that lagged dependence can be accommodated within
our framework, treating the case $\nu=2$.

\subsection{Regular marked point process and  loglinear modeling}

In this subsection we prove that the heuristic arguments of Section
\ref{sec2}
for marginal methods hold under mild conditions.
 Consider $\nu \geq
1$ neurons labeled $1$ to~$\nu$. For $T>0$, let $N_T$ denote the total
number of spikes produced by these $\nu$ neurons on the time interval
$[0, T)$ and let $0\leq s_1 <\cdots < s_{N_T} <T$ denote the specific
spike times. For each $j=1,\ldots, N_T$, we write $(s_j, (i_j) )$ to
represent the event that a spike was fired at time $s_j$ by neuron
$i_j$ where $i_j \in \{1, \ldots, \nu\}$.  We observe from
Section \ref{sec:process} that
\[
{\mathcal H}_T = \{ (s_1, (i_1) ), \ldots, (s_{N_T}, (i_{N_T}) )\}
\]
forms a marked point process on the interval $[0, T)$ with mark space
${\mathcal K} = \{(1), \ldots, (\nu)\}$. Following the notation of
Section \ref{sec:process}, let $\lambda (t, (i)| {\mathcal H}_t)$ denote
the conditional intensity function of the point process ${\mathcal H}_T$.
We assume that the following two conditions hold:

\textit{Condition} (I). There exists a strictly positive refractory period
for each neuron in that there exists  a constant $\theta>0$ such that
$\lambda( t, (i)| {\mathcal H}_t) = 0$ if there exists some $(s_, (i)) \in
{\mathcal H}_t$ such that $t-s \leq \theta, i \in \{1,\ldots, \nu\}$.

\textit{Condition} (II). For each $k\in \{0,\ldots, 2 \lceil T/\theta\rceil
-1\}$ and $i, i_1,\ldots, i_k\in \{1,\ldots, \nu\}$, the conditional
intensity function $\lambda (t, (i) | \{ (s_1, (i_1) ),\ldots, (s_k,
(i_k) ) \} )$ is a continuously differentiable function
 in $(s_1,\ldots, s_k, t)$ over the simplex $0\leq s_1 \leq \cdots \leq s_k \leq  t \leq  T$.

If $\delta <\theta$, then condition (I) implies that there is at most 1
spike from each neuron in a bin of width $\delta$. Conditions (I) and
(II) also imply that the marked point process is regular in that
(exactly) synchronous spikes occur only with probability~0. Theorem
\ref{tm:4.3} below gives the limiting relationship between the bin
probabilities of the induced discrete-time process and the conditional
intensities of the underlying continuous-time marked point process.

\begin{tm} \label{tm:4.3}
Suppose that conditions \textup{(I)} and \textup{(II)} hold, $1\leq i_1<\cdots < i_k \leq
\nu$ and $1\leq k\leq \nu$. Then
\begin{eqnarray*}
&& \lim_{\delta \rightarrow 0} \delta^{-k} P^{i_1,\ldots, i_k}_{1,
\ldots, 1} ( t_m)
\nonumber \\
&&\qquad= \frac{1}{k!} \sum_{j_1, \ldots, j_k : \{j_1,\ldots, j_k\} =
\{i_1,\ldots, i_k\} }
 E  \prod_{l =1}^k \bigl[ \lambda \bigl( t, (j_k)| \{ (t, (j_1))\} \\
 &&\qquad\hphantom{= \frac{1}{k!} \sum_{j_1, \ldots, j_k : \{j_1,\ldots, j_k\} =
\{i_1,\ldots, i_k\} }
 E  \prod_{l =1}^k \bigl[ \lambda \bigl(}\cup \cdots \cup \{ (t, (j_{k-1}))\} \cup {\mathcal H}_t \bigr)\bigr],
\end{eqnarray*}
where $t_m =m\delta \rightarrow t$ as $\delta \rightarrow 0$ and
$\lambda( t, (i_2)| \{ (t, (i_1))\} \cup {\mathcal H}_t) = \lim_{t^*
\rightarrow t-} \lambda( t,\break (i_2)| \{ (t^*, (i_1))\} \cup {\mathcal H}_t)$,
etc. Here the expectation is taken with respect to ${\mathcal
H}_t$.
\end{tm}

Theorem \ref{tm:4.3} validates the heuristics stated in (\ref{eq:2.2})
where $\nu=2$,
\begin{eqnarray*}
\lambda^i (t) &=&  E \bigl[ \lambda \bigl(t, (i)| {\mathcal H}_t\bigr) \bigr],\\
\lambda^{1, 2} (t) &=& \frac{1}{2} \sum_{1\leq i_1\neq i_2\leq 2} E
\bigl[ \lambda \bigl( t, (i_2)| \{ (t, (i_1))\} \cup {\mathcal H}_t \bigr)
\lambda (t, (i_1)| {\mathcal H}_t ) \bigr].
\end{eqnarray*}

Next we construct the discrete-time loglinear model induced by the
above marked point process. First define recursively for $t_m=m\delta$,
\begin{eqnarray} \label{eq:4.90}
\zeta_{\{i_1\}} (t_m) &=& \delta^{-1} P^{i_1}_1
(t_m)\qquad\forall i_1=1, \ldots, \nu,\nonumber \\
\zeta_{\{i_1, i_2\}} (t_m) &=& \frac{ \delta^{-2} P^{i_1, i_2}_{1,1}
(t_m) }{ \zeta_{\{i_1\}} (t_m) \zeta_{\{i_2\}} (t_m)}\qquad
\forall 1\leq i_1< i_2 \leq \nu,\nonumber \\ [-8pt]\\ [-8pt]
\vdots &&\nonumber \\
\zeta_{\{i_1, \ldots, i_k\}} (t_m) &=& \frac{ \delta^{-k}
P^{i_1,\ldots, i_k}_{1,\ldots, 1} (t_m) }{ \prod_{\Xi\subsetneq
\{i_1,\ldots, i_k\}} \zeta_{\Xi} (t_m) }\nonumber\\
\eqntext{\forall 1\leq
i_1< \cdots <i_k \leq \nu, 2\leq k\leq \nu.}
\end{eqnarray}
We further define
\begin{equation}
\hspace*{15pt}\xi_{\{i_1, \ldots, i_k\}} (t) = \lim_{\delta \rightarrow 0}
\zeta_{\{i_1, \ldots, i_k\}} (t_m)\qquad\forall 1\leq i_1<
\cdots <i_k \leq \nu, 1\leq k\leq \nu, \label{eq:4.57}
\end{equation}
where $\lim_{\delta\rightarrow 0} t_m \rightarrow t$, whenever the
expression on the right-hand side of (\ref{eq:4.57}) is well defined.
The following is an immediate corollary of Theorem \ref{tm:4.3}.

\begin{cy} \label{cy:4.5}
Let $\xi_{\{i_1\}} (t)$ and  $\xi_{\{i_1, \ldots, i_k\}}(t)$ be as in
(\ref{eq:4.57}). Then with the notation and assumptions of Theorem
\ref{tm:4.3}, we have
\begin{eqnarray*}
\xi_{\{i_1\}} (t) &=& E [\lambda (t, (i_1)| {\mathcal H}_t )],\\
\xi_{\{i_1, \ldots, i_k\}} (t) &=& \biggl[ k!  \prod_{\Xi\subsetneq
\{i_1,\ldots, i_k\}} \zeta_\Xi (t)  \biggr]^{-1}\\
&&{}\times \sum_{\{j_1,\ldots, j_k\} = \{i_1,\ldots, i_k\} }
 E  \prod_{l =1}^k \bigl[ \lambda \bigl( t, (j_k)| \{ (t, (j_1))\}\\
 &&\hphantom{{}\times \sum_{\{j_1,\ldots, j_k\} = \{i_1,\ldots, i_k\} }
 E  \prod_{l =1}^k \bigl[ \lambda \bigl(} \cup \cdots \cup \{ (t, (j_{k-1}))\} \cup {\mathcal H}_t \bigr)
 \bigr],
\end{eqnarray*}
whenever the right-hand sides are well defined.
\end{cy}

It is convenient to define $\tilde{P}^{1,\ldots, \nu}_{0,\ldots, 0}
(t_m) = 1$. For $a_1,\ldots, a_\nu \in \{0,1\}$ and not all~0, define
$\tilde{P}^{1,\ldots, \nu}_{a_1,\ldots, a_\nu} (t_m) = P^{i_1,\ldots,
i_k}_{1,\ldots, 1} (t_m)$
where $i\in \{ i_1,\ldots, i_k\}$ if and only if $a_i= 1$. Then using
the notation of (\ref{eq:4.90}), the corresponding loglinear model
induced by the above marked point process is
\begin{eqnarray}\label{eq:4.91}
&&\log [  \tilde{P}^{1, \ldots, \nu}_{a_1, \ldots, a_\nu} (t_m) ]\nonumber\\
&&\qquad=\log [ P^{i_1, \ldots, i_k}_{1, \ldots, 1} (t_m) ]\\
&&\qquad= \sum_{i=1}^\nu a_i \log [ P^i_1 (t_m) ] + \sum_{\Xi\subseteq
\{1,\ldots, \nu\}: |\Xi| \geq 2}  \biggl(\prod_{j\in \Xi} a_j\biggr) \log[
\zeta_\Xi (t_m) ]\nonumber
\end{eqnarray}
for all $a_1, \ldots, a_\nu \in \{0,1\}$.  Under conditions (I) and
(II), Corollary \ref{cy:4.5} shows that $\xi_\Xi (t) =
\lim_{\delta\rightarrow 0} \zeta_\Xi (t_m)$ is continuously
differentiable.  This gives an asymptotic justification for smoothing
the estimates of $\zeta_\Xi$, $\Xi\subseteq \{1,\ldots, \nu\}$.

\subsection{Case of $\nu=2$ neurons with lag $h$}

This subsection considers the lag $h$ case for two neurons labeled 1
and 2. Let $h, m$ be integers such that $0\leq m \leq m+h \leq  T
\delta^{-1} -1$.  As in (\ref{eq:1.35}), we write
\[
P^{1, 2}_{a, b} (t_m, t_{m+h}) = P [X^2 (t_{m+h}) = b, X^1(t_m) =a ]\qquad
\forall a, b\in \{0,1\},
\]
where $t_m= m\delta$ and $t_{m+h}=(m+h)\delta$. Analogous to Theorem
\ref{tm:4.3}, we have the following results for the lag case.

\begin{tm} \label{tm:4.2}
Suppose conditions \textup{(I)} and \textup{(II)} hold. Then
\[
\lim_{\delta \rightarrow 0} \frac{ P^{1,2}_{1,1} (t_m, t_{m+h})
}{\delta^2} =  E \bigl[ \lambda \bigl( t+\tau, (2)| \{ (t, (1))\} \cup
{\mathcal H}_{t+\tau} \bigr) \lambda (t, (1)| {\mathcal H}_t ) \bigr],
\]
where $t_{m+h}\rightarrow t+\tau$ and $t_m\rightarrow t$ as $\delta
\rightarrow 0$ for some constant $\tau>0$. Here the expectation is
taken with respect to ${\mathcal H}_{t+\tau}$ (and hence also ${\mathcal
H}_t$).
\end{tm}

\begin{cy}
Let $\zeta(t_m, t_{m+h})$ be defined as in (\ref{eq:2.42}). Then with
the notation and assumptions of Theorem \ref{tm:4.2}, we have
\begin{eqnarray}\label{eq:4.78}
&&\hspace*{20pt}\lim_{\delta \rightarrow 0} \zeta (t_m, t_{m+h})\nonumber \\ [-8pt]\\ [-8pt]
&&\hspace*{20pt}\qquad= \frac{ E[ \lambda(
t+\tau, (2)| \{ (t, (1))\} \cup {\mathcal H}_{t+\tau}) \lambda (t, (1)|
{\mathcal H}_t )] }{ E[ \lambda (t+\tau, (2)| {\mathcal H}_{t+\tau})] E[
\lambda( t, (1)|{\mathcal H}_t)] }\qquad \forall 0\leq t <T-\tau,\nonumber
\end{eqnarray}
whenever the right-hand side is well defined.
\end{cy}

We observe from conditions (I) and (II) that the right-hand side of
(\ref{eq:4.78}) is continuously differentiable in $t$. Again this
provides an asymptotic justification for smoothing the estimate of
$\zeta(t, t+\tau)$, with respect to $t$, when $\delta$ is small.

\section{Theoretical results: Conditional methods}\label{sec:conditional}

This section is analogous to Section \ref{sec:marginal}, but treats the
conditional case. We (i) provide a justification of the limiting
statements in (\ref{eq:2.83}) and (ii) generalize to higher-order
models. We again also note that lagged dependence can be accommodated
within our framework, treating the case $\nu=2$.

\subsection{Synchrony and loglinear modeling}\label{sec61}

This subsection considers $\nu \geq 1$ neurons labeled $1$ to $\nu$. We
model the spike trains generated by these neurons on $[0, T)$ by a
marked point process ${\mathcal H}_T$ with mark space
\[
{\mathcal K}= \{ (i_1,\ldots, i_k)\dvtx 1\leq i_1<\cdots < i_k\leq \nu,
k=1,\ldots, \nu \}.
\]
Here, for example, the mark $(i_1)$ denotes the event that neuron $i_1$
(and only this neuron) spikes, $(i_1, i_2)$ denotes the event that
neuron $i_1$ and neuron $i_2$ (and only these two neurons) spike in
synchrony (i.e., at the same time), and the mark $(1, \ldots, \nu)$
denotes the event that all $\nu$ neurons spike in synchrony.

Let $N_t$ denote the total number of spikes produced by these neurons
on $[0,t)$, $0<t\leq T$, and let $0\leq s_1< \cdots < s_{N_T} <T$
denote the specific spike times. For each $j=1,\ldots, N_T$, let
$\kappa_j \in {\mathcal K}$ be the mark associated with $s_j$. Then ${\mathcal
H}_T$ can be expressed as
\begin{equation}\label{eq:5.5}
{\mathcal H}_T = \{ (s_1,\kappa_1), \ldots, (s_{N_T}, \kappa_{N_T})\}.
\end{equation}

Given ${\mathcal H}_t$, we write
\begin{eqnarray}
{\mathcal H}^i_t = \{ s\dvtx (s,\kappa) \in {\mathcal H}_t \mbox{ for some }
\kappa = ( i_1, \ldots, i_k) \mbox{ such that } i\in \{i_1,\ldots, i_k\}\}\nonumber\\
\eqntext{\forall i=1,\ldots, \nu.}
\end{eqnarray}
${\mathcal H}^i_t$ denotes the spiking history of neuron $i$ on $[0, t)$.
The conditional intensity function $\lambda(t, \kappa| {\mathcal H}_t)$,
$t\in [0, T)$ and $\kappa\in {\mathcal K}$, of the marked point process
${\mathcal H}_T$ is defined to be
\begin{eqnarray}\label{eq:5.6}
\lambda (t, (i) | {\mathcal H}_t ) &=& \lambda^i (t| {\mathcal H}^i_t)\qquad\forall t\in [0,T),\nonumber
\\ [-8pt]\\ [-8pt]
\lambda( t, (i_1, \ldots, i_k)| {\mathcal H}_t) &=& \delta^{k-1}
\gamma_{\{i_1, \ldots, i_k\}} (t) \prod_{j=1}^k \lambda^{i_j} (t| {\mathcal
H}^{i_j}_t)\qquad\forall
 t\in [0, T),\nonumber
\end{eqnarray}
where $\delta>0$ is a constant, $\gamma_{\{i_1, \ldots, i_k\}}(t)$'s
are functions depending only on $t$ and the $\lambda^i( t|{\mathcal
H}^i_t)$'s are conditional intensity functions depending only on the
spiking history of neuron $i$. We take $\gamma_{\{i\}} (t)$ to be
identically equal to 1.

From (\ref{eq:5.6}), we note that the above marked point process model
is not a~single marked point process but rather a family of marked
point processes indexed by $\delta$. In the sequel, we let
$\delta\rightarrow 0$. We further assume that the following two
conditions hold:

\textit{Condition} (III). There exists a strictly positive refractory
period for each neuron in that there is a constant $\theta>0$ such
that, for $i=1, \ldots, \nu$ and $t\in [0, T)$,
\[
\lambda^i ( t | {\mathcal H}^i_t) = 0,\qquad\mbox{if there exists
some } s \in {\mathcal H}^i_t \mbox{ such that }  t-s \leq \theta.
\]

\textit{Condition} (IV). For each $k\in \{0,\ldots, \lceil T/\theta \rceil
-1\}$ and  $i\in \{1, \ldots, \nu\}$, $\lambda^i (t | \{s_1,\break\ldots, s_k
\} )$ is a continuously differentiable function
 in $(s_1,\ldots, s_k, t)$ over the simplex $0\leq s_1 \leq \cdots \leq s_k \leq  t \leq  T$.

Following Section \ref{sec22}, we divide the time interval $[0, T)$ into bins
of width~$\delta$. For simplicity, we assume that $T$ is a multiple of
$\delta$. Let $t_m=m\delta$ and $X^i(t_m)$, $m= 0,\ldots, T \delta^{-1}
-1$, be as in Section \ref{sec1}. If $X^i (t_l)=1$ for all $l\in \{ l_1,\ldots,
l_k\}$, and $X^i(t_l)=0$ otherwise, for some subset $0\leq l_1 < \cdots
< l_k \leq m-1$, we write
\begin{equation}\label{eq:5.34}
\bar{\mathcal H}_{t_m}^i = \{ t_{l_1},\ldots, t_{l_k} \}, \qquad
\bar{\mathcal H}_{t_m} = (\bar{\mathcal H}_{t_m}^1, \ldots, \bar{\mathcal
H}_{t_m}^{\nu}).
\end{equation}
It should be observed that although the above definitions of ${\mathcal
H}_{t_m}^i$ and $\bar{\mathcal H}_{t_m}$ differ from those given in Section
\ref{sec:overview}, they are equivalent. We note that the conditional intensity
functions $\lambda (t, (i_1, \ldots, i_k)|{\mathcal H}_t)$ in
(\ref{eq:5.6}) depend on the bin width $\delta$. This is necessary in
order to preserve the  natural hierarchical sparsity conditions given
by
\[
\sup_{m= 0,\ldots, T \delta^{-1} -1} P [ X^{i_1} (t_m) = 1, \ldots,
X^{i_k}(t_m)=1 ] = O(\delta^k),
\]
as $\delta \rightarrow 0$ for all $1\leq i_1< \cdots< i_k\leq \nu$,
$1\leq k\leq \nu$.

\begin{tm} \label{tm:5.3}
Consider the marked point process ${\mathcal H}_T$ as in (\ref{eq:5.5})
with conditional intensity function satisfying (\ref{eq:5.6}). Then
under conditions \textup{(III)} and \textup{(IV)}, we have
\begin{eqnarray*}
P^i_1 (t_m| \bar{\mathcal H}_{t_m}) &=& \delta \lambda^i (t_m | \bar{\mathcal
H}_{t_m}^i ) + O(\delta^2),\nonumber \\
P^{i_1, i_2}_{1, 1} (t_m | \bar{\mathcal H}_{t_m} ) &=& \delta^2 \bigl[ 1 +
\gamma_{\{i_1, i_2\}} (t_m) \bigr] \prod_{j=1}^2 [ \lambda^{i_j} (t_m |
\bar{\mathcal H}_{t_m}^{i_j} ) ] +O(\delta^3),
\end{eqnarray*}
and in general,
\begin{eqnarray*}
&& P^{i_1, \ldots, i_k}_{1, \ldots, 1} (t_m | \bar{\mathcal H}_{t_m} )\nonumber \\
&&\qquad= \delta^k \Biggl\{ \sum_{j=1}^k \sum_{\Xi_1, \ldots, \Xi_j: \mathrm{
all\ disjoint\ and\ nonempty}, \cup \Xi_j = \{i_1,\ldots, i_k\}}
\prod_{l=1}^j \gamma_{\Xi_l} (t_m ) \Biggr\}\\
&&\qquad\quad{}\times \prod_{\ell =1}^k [
\lambda^{i_\ell} (t_m | \bar{\mathcal H}_{t_m}^{i_\ell} ) ]
+O(\delta^{k+1})
\end{eqnarray*}
for sufficiently small $\delta$ where $\bar{\mathcal H}_{t_m}$ and
$\bar{\mathcal H}^i_{t_m}$ are defined by (\ref{eq:5.34}).
\end{tm}

The following is an immediate corollary of Theorem \ref{tm:5.3}. It
gives an asymptotic justification for equation (\ref{eq:2.8}) in
Section \ref{sec22}.
\begin{cy} \label{cy:5.1}
With the notation and assumptions of Theorem \ref{tm:5.3}, we have for
$\nu=2$,
\[
 P^{1, 2}_{1, 1} (t_m | \bar{\mathcal H}_{t_m} )
 = \zeta (t_m) P^1_1 (t_m| \bar{\mathcal H}^1_{t_m} ) P^2_1 (t_m| \bar{\mathcal H}^2_{t_m} ) +O(\delta^3)
\]
for sufficiently small $\delta> 0$ uniformly over $\bar{\mathcal H}_{
t_m}^i$, $m=0,\ldots, T \delta^{-1} -1$ where $\zeta(t_m) = 1 +
\gamma_{\{1,2\}} (t_m)$.
\end{cy}

We now use Theorem \ref{tm:5.3} to construct a loglinear model (for the
above spike train data) whose higher-order coefficients are
asymptotically independent of past spiking history.  First define
recursively
\begin{eqnarray}\label{eq:5.7}
\zeta_{\{i_1\}} (t_m | \bar{\mathcal H}_{t_m}) &=& \delta^{-1} P^{i_1}_1
(t_m | \bar{\mathcal H}_{t_m})\qquad\forall i_1=1,\ldots, \nu,
\nonumber \\
\zeta_{\{i_1, i_2\}} (t_m | \bar{\mathcal H}_{t_m}) &=& \frac{ \delta^{-2}
P^{i_1, i_2}_{1,1} (t_m | \bar{\mathcal H}_{t_m}) }{ \zeta_{\{i_1\}} (t_m |
\bar{\mathcal H}_{t_m}) \zeta_{\{i_2\}} (t_m | \bar{\mathcal H}_{t_m})}\qquad\forall 1\leq i_1< i_2 \leq \nu,
\nonumber \\ [-8pt]\\ [-8pt]
\vdots && \nonumber \\
\zeta_{\{i_1, \ldots, i_k\}} (t_m | \bar{\mathcal H}_{t_m}) &=& \frac{
\delta^{-k} P^{i_1, \ldots, i_k}_{1, \ldots, 1} (t_m | \bar{\mathcal
H}_{t_m})  }{ \prod_{\Xi \subsetneq \{i_1,\ldots, i_k\}} \zeta_{\Xi}
(t_m | \bar{\mathcal H}_{t_m}) }\qquad \forall 1\leq i_1<\cdots
<i_k\leq \nu.\nonumber
\end{eqnarray}

It follows from Theorem \ref{tm:5.3} and (\ref{eq:5.7}) that for
sufficiently small $\delta$,
\begin{eqnarray*}
&&\hspace*{9.1pt}\zeta_{\{i_1\}} (t_m | \bar{\mathcal H}_{t_m}) = \lambda^{i_1} (t_m |
\bar{\mathcal H}_{t_m}^{i_1} ) + O(\delta)\qquad\forall i_1=1,
\ldots, \nu,\\
&&\zeta_{\{i_1, i_2\}} (t_m | \bar{\mathcal H}_{t_m}) = 1 + \gamma_{\{i_1,
i_2\}} (t_m) + O(\delta)\qquad\forall 1\leq i_1< i_2 \leq
\nu,\\
&&\hspace*{62.5pt}\vdots  \\
&&\zeta_{\{i_1, \ldots, i_k\} } (t_m | \bar{\mathcal H}_{t_m})\\
&&\qquad= \frac{
\sum_{j=1}^k \sum_{\Xi_1, \ldots, \Xi_j: \mbox{\tiny all disjoint and
nonempty}, \cup \Xi_j = \{i_1,\ldots, i_k\}} \prod_{l=1}^j
\gamma_{\Xi_l} (t_m ) }{ \prod_{\Xi \subsetneq \{i_1,\ldots, i_k\}:
|\Xi|\geq 2} \zeta_{\Xi} (t_m | \bar{\mathcal H}_{t_m})
 } + O(\delta)
\end{eqnarray*}
whenever $1\leq i_1<\cdots <i_k\leq \nu$ and $k\geq 2$, assuming that
terms on the right-hand side are well defined. The practical importance
of these results lies in the fact that the coefficients $\zeta_{\{i_1,
\ldots, i_k\}} (t_m | \bar{\mathcal H}_{t_m})$ with $k\geq 2$ are
asymptotically (as $\delta\rightarrow 0$) independent of $\bar{\mathcal
H}_{t_m}$, the spiking history of the\vspace*{-1pt} neurons. It is convenient to
define $\tilde{P}^{1, \ldots, \nu}_{0, \ldots, 0} (t_m  | \bar{\mathcal
H}_{t_m}) = 1$. For $a_1,\ldots, a_\nu \in \{0,1\}$ and not all~0,
define $\tilde{P}^{1, \ldots, \nu}_{a_1, \ldots, a_\nu} (t_m |
\bar{\mathcal H}_{t_m}) = P^{i_1,\ldots, i_k}_{1,\ldots, 1} (t_m |
\bar{\mathcal H}_{t_m})$ where $i \in \{i_1,\ldots, i_k\}$ if and only if
$a_i=1$. Then the induced loglinear model is
\begin{eqnarray}\label{eq:5.76}
&& \hspace*{20pt}\log [ \tilde{P}^{1, \ldots, \nu}_{a_1, \ldots, a_\nu} (t_m |
\bar{\mathcal H}_{t_m}) ]\nonumber \\ [-8pt]\\ [-8pt]
&&\hspace*{20pt}\qquad= \sum_{i=1}^\nu a_i \log[ P^i_1 (t_m | \bar{\mathcal H}_{t_m}) ] +
\sum_{\Xi \subseteq \{1,\ldots, \nu\}: |\Xi| \geq 2} \biggl(\prod_{j\in \Xi}
a_j\biggr) \log [ \zeta_\Xi ( t_m | \bar{\mathcal H}_{t_m}) ]\nonumber
\end{eqnarray}
for all $a_1, \ldots, a_\nu \in \{0,1\}$ where the second term on the
right-hand side of~(\ref{eq:5.76}) is asymptotically (as
$\delta\rightarrow 0$) independent of the spiking history $\bar{\mathcal
H}_{t_m}$.

\subsection{$\nu=2$ neurons with time-delayed synchrony}

This subsection considers $\nu=2$ neurons labeled 1, 2 and let $\tau
>0$ be a constant denoting the spike lag. We model the spike train
generated by the two neurons on $[0, T)$ by a~marked point process
${\mathcal H}_T$ as in (\ref{eq:8.1}) with mark space ${\mathcal K}= \{ (1),
(2), (1, 2)\}$. The marks $(1), (2)$ are interpreted as before as
isolated (i.e., nonsynchronous) spikes. However, now $(1,2)$ is
interpreted to be neuron 1 spiking first and then neuron 2 spiking
second after a delay of $\tau$ time units. The mark $(1,2)$ is used to
model a precise time-delayed synchronous spiking of lag $\tau$ between
the 2 neurons.

Let $N_T$ denote the number of times the three marks occur on $[0,T)$
and $s_1< \cdots < s_{N_T}$ be the specific spike times. For each
$j=1,\ldots, N_T$, let $\kappa_j \in {\mathcal K}$ be the mark associated
with $s_j$. Then ${\mathcal H}_t$ can be decomposed into $({\mathcal H}_t^1,
{\mathcal H}_{t+\tau}^2)$, where
\begin{eqnarray}\label{eq:5.3}
{\mathcal H}_t^1 &=& \bigl\{ s\dvtx (s, \kappa)\in {\mathcal H}_t \mbox{ where }
\kappa\in \{(1), (1,2)\} \bigr\},\nonumber \\ [-8pt]\\ [-8pt]
{\mathcal H}_{t+\tau}^2 &=& \bigl\{ s\dvtx (s, \kappa)\in {\mathcal
H}_{t+\tau} \mbox{where} \kappa\in \{(2), (1,2)\} \bigr\}.\nonumber
\end{eqnarray}
To be definite, $(s, \kappa) = (s, (1,2))$ means neuron 1 spikes at
time $s$ and neuron~2 spikes at time $s + \tau$. The conditional
intensity function $\lambda(t, \kappa| {\mathcal H}_t)$, $t\in [0, T)$ and
$\kappa\in {\mathcal K}$, of the marked point process ${\mathcal H}_T$ is
defined to be
\begin{eqnarray}\label{eq:5.4}
\lambda (t, (i) | {\mathcal H}_t ) &=& \lambda^i (t| {\mathcal H}^i_t)\qquad\forall i=1,2,\ t\in [0,T),
\nonumber \\ [-8pt]\\ [-8pt]
\lambda (t, (1,2) | {\mathcal H}_t) &=& \delta \gamma (t, t+\tau)
\lambda^1(t | {\mathcal H}^1_t) \lambda^2(t + \tau | {\mathcal H}_{t+\tau}^2 )\qquad\forall t\in [0, T),\nonumber
\end{eqnarray}
where $\delta>0$ is a constant, $\gamma(t, t + \tau)$ is a continuously
differentiable function in~$t$ on the interval $0\leq t \leq T - \tau$,
and $\lambda^1( t|{\mathcal H}^1_t)$,
 $\lambda^2(t + \tau |  {\mathcal H}_{t+\tau}^2 )$
are  conditional intensity functions depending only on the spiking
history of neuron $1$ up to time $t$ and on the spiking history of
neuron $2$ up to time $t+\tau$, respectively.

As in Section \ref{sec1}, we divide the time interval $[0, T)$ into bins of
width $\delta >0$. Let $\bar{\mathcal H}_{t_m} = ( \bar{\mathcal H}^1_{t_m},
\bar{\mathcal H}^2_{t_{m+h}})$ be as in (\ref{eq:5.34}) where, for
simplicity, we assume that~$\delta$ is chosen such that $m, h$ are
integers satisfying $0\leq m\leq m+h\leq T \delta^{-1} -1$ and $t_{m+h}
= t_m + \tau$. Recall that, by definition,
\begin{eqnarray*}
P^i_1( t_m | \bar{\mathcal H}_{t_m}^i ) &=& P\bigl(X^i (t_m) =1 | \bar{\mathcal
H}_{t_m}^i \bigr),\\
P^{1,2}_{1,1} (t_m, t_{m+h}| \bar{\mathcal H}_{t_m} ) &=& P\bigl( X^1 (t_m) =1,
X^2(t_{m+h}) =1 | \bar{\mathcal H}_{t_m} \bigr).
\end{eqnarray*}

\begin{tm} \label{tm:5.2}
Consider the marked point process ${\mathcal H}_T$ as in (\ref{eq:5.3})
with conditional intensity function satisfying (\ref{eq:5.4}). Let $m,
h$ be integers satisfying $0\leq m\leq m+h\leq T \delta^{-1} -1$ and
$t_{m+h} = t_m + \tau$. Then under conditions \textup{(III)} and \textup{(IV)}, we have
\[
P^{1, 2}_{1, 1} (t_m, t_{m+h} | \bar{\mathcal H}_{t_m} ) = \delta^2 [
\gamma (t_m, t_{m+h}) +1] P^1_1( t_m | \bar{\mathcal H}_{t_m}^1 ) P^2_1(
t_{m+h} | \bar{\mathcal H}_{t_{m+h}}^2 )
+O(\delta^3)
\]
for sufficiently small $\delta$.
\end{tm}

The practical significance of Theorem \ref{tm:5.2} is that $\gamma(t_m,
t_{m+h})$ does not depend on the spiking history of the 2 neurons.
This implies that a statistic based on $\gamma$ can be constructed to
test the null hypothesis $H_0$ that there is no time-delayed spiking
synchrony at lag $\tau$ between the 2 neurons.

\section{Discussion}\label{sec7}

We have described an approach to assessing spike train synchrony using
loglinear models for multiple binary time series. We tried to motivate
the application of loglinear modeling technology in Section \ref{sec1},
emphasizing two features of individual neural response:
stimulus-induced nonstationarity that remains time-locked across
trials, and within-trial effects that are history-dependent, with
timing that varies across trials. These were incorporated into the
models by including for individual neurons both time-varying marginal
effects, which stay the same across trials, and history-dependent
terms; interaction terms were treated separately. In
Section \ref{sec:analysis} we presented results for two pairs of
neurons. For both pairs there was evidence of excess synchronous
spiking beyond that explained by stimulus-induced changes in
individual-neuron firing rates. In one pair, network activity,
represented as history dependence, was sufficient to account for excess
synchronous spiking, but the other pair displayed excess synchronous
spiking that remained highly statistically significant even after
network effects were incorporated, indicating stimulus-related
synchrony. Our theoretical results provided a~continuous-time point
process foundation for the methods, justifying both our use of
smoothing and our derivation of the excess-synchrony estimators $\hat
\zeta$ and $\hat \zeta_H$.

Assessment of synchrony via continuous-time loglinear models is closely
related to the unitary-event analysis of Gr\"un, Diesmann and Aertsen
(\citeyear{GDA2002a}, \citeyear{GDA2002b}). Unitary event analysis assumes each neuron follows a
locally-statio\-nary Poisson process, which has been shown to be somewhat
conservative in the sense of providing inflated $p$-values in the
presence of non-Poisson history dependence. Its main purpose is to
identify stimulus-locked excess synchrony. Because the loglinear models
could be viewed as generalizations of locally-stationary Poisson
models, they could extend unitary-event analysis to cases in which it
seems desirable to account more explicitly for stimulus and history
effects. This is a topic for future research.

We also provided an example of testing for 3-way interaction. The
results we gave in Section \ref{sec:analysis} for a particular triple
of neurons indicated no evidence of excess 3-way joint spiking above
that explained by 2-way joint spiking. A systematic finding along these
lines, examining large numbers of neurons, would be consistent with
findings of Schneidman et al. (\citeyear{SBSB2006}). However, as may be seen from
Figure~\ref{fig3}, 3-way joint spikes are very sparse. A careful study
of the power to detect 3-way joint spiking in contexts like the one
considered here could be quite helpful. We plan to carry out such a
study and report it elsewhere.

We have restricted history effects to individual neurons by assuming,
first, that each neuron's history excludes past spiking of the other
neurons under consideration and, second, that the interaction effects
are independent of history. This greatly simplifies the modeling and
avoids confounding the interaction effects with cross-neuron effects.
While it would be possible, in principle (by modifying the hierarchical
sparsity condition), to allow history-dependence within interaction
effects, we see no practical benefit of doing so. With the two
additional, highly plausible assumptions used here, we get both
tractable discrete-time methods and a sense in which the methods may be
understood in continuous time. A key element of our formalism is the requirement of hierarchical
sparsity, as in the special case of equation~(\ref{eq:sparse2}) and more generally in Section
\ref{sec61} (preceding Theorem \ref{tm:5.3}). This corresponds to the practical reality
that two-way synchronous spikes are rare, as in Figure~\ref{fig2}, and three-way spikes
are even rarer, as in Figure \ref{fig3}. Some form of sparsity seems to us essential.
[After our article was accepted we became aware that Solo
(\citeyear{Solo2007}) had attempted to develop likelihoods for point processes having
synchronous events, but because his approach does not incorporate
sparsity we have been unable to understand how it could be used in
the kind of applications we have described here.] It is somewhat inelegant to have a
sequence of marked processes (indexed by $\delta$), but this appears to
be the best that can be achieved by starting with very natural
discrete-time loglinear models. An alternative would be to use more
standard point process models with short time-scale cross-neuron
effects. Presumably, similar results could be obtained, but the
relationship between these different approaches is also a subject for
future research. A quite different technology involves
permutation-style assessment via ``dithering'' or ``jittering'' of
individual spike times [cf. Geman et al. (\citeyear{GAHH2010}), Gr\"un (\citeyear{Grun2009})].
Synchrony is one of the deep topics in computational neuroscience and
its statistical identification is subtle for many reasons, including
inaccuracies in reconstruction of spike timing from the complicated
mixture of neural signals picked up by the recording electrodes [e.g.,
Harris et al. (\citeyear{HHCHB2000}), Ventura (\citeyear{Ventura2009})]. It is likely that multiple
approaches will be needed to grapple with varying neurophsyiological
circumstances.

\begin{appendix}
\section*{Appendix}\label{appendix}

\begin{pf*}{Proof of Theorem \ref{tm:4.3}}
For simplicity, we shall consider
only the case for $\nu=2$. The proof for other values of $\nu$ is
similar though more tedious. We observe from Proposition \ref{pn:8.1}
that
\begin{eqnarray}\label{eq:a.10}
&&\frac{P^{1, 2}_{1, 0} (t_m)  }{\delta}\nonumber \\
&&\qquad= \frac{1}{\delta} \sum_{k=
0}^{2 \lceil T/\theta\rceil}  \sum_{i_1,\ldots, i_k\in \{1,2\}}
\int_0^{t_m} \ldots \int_{s_{k-1}}^{t_m}\int_{t_m}^{t_{m+1}}\nonumber\\
&&\qquad\quad\hspace*{3.3pt}\lambda
(s_{k+1}, (1)| \{ (s_1, (i_1) ), \ldots, (s_k, (i_k) ) \} ) \\
&&\qquad\quad{} \times \Biggl[ \prod_{l=1}^k \lambda (s_l, (i_l) | \{ (s_1,
(i_1) ), \ldots, (s_{l-1}, (i_{l-1}) ) \} ) \Biggr]
\nonumber \\
&& \qquad\quad{} \times e^{-\sum_{j=1}^2 \int_0^{t_{m+1}} \lambda (w,
(j) | {\mathcal H}_w)\,dw}\,ds_{k+1}\,ds_k \cdots\,ds_1.\nonumber
\end{eqnarray}
Condition (I) implies that the summations $\sum_k  \sum_{i_1,\ldots,
i_k\in \{1,2\}}$ in (\ref{eq:a.10}) contain a finite number of
summands. Hence, letting $\delta\rightarrow 0$, the right-hand side of~%
(\ref{eq:a.10}) equals
\begin{eqnarray}\label{eq:a.20}
&& \sum_{k= 0}^{2 \lceil T/\theta\rceil}\!\sum_{i_1,\ldots, i_k\in
\{1,2\}}\!\lim_{\delta \rightarrow 0} \frac{1}{\delta}\!\int_0^{t_m}
\!\!\!\!\ldots\!\int_{s_{k-1}}^{t_m}\!\int_{t_m}^{t_{m+1}}\!\lambda (\!s_{k+1}, (1)|
\{\! (\!s_1, (i_1)\!), \ldots, (\!s_k, (i_k)\!)\!\}\!)\hspace*{-12pt}\nonumber\\
&&\qquad{} \times \Biggl[ \prod_{l=1}^k \lambda (s_l, (i_l) | \{ (s_1,
(i_1) ), \ldots, (s_{l-1}, (i_{l-1}) ) \} ) \Biggr]\\
&&\qquad{}\times e^{-\sum_{j=1}^2 \int_0^{t_{m+1}} \lambda (w,
(j) | {\mathcal H}_w)\,dw}\,ds_{k+1}\,ds_k \cdots\,ds_1.\nonumber
\end{eqnarray}
Using Condition (II) and the Taylor expansion, we have
\begin{eqnarray*}
&& \lambda (s_{k+1}, (1)| \{ (s_1, (i_1) ), \ldots, (s_k, (i_k) ) \} )
\nonumber \\
&&\qquad= \lambda ( t_m, (1)| \{ (s_1, (i_1) ), \ldots, (s_k, (i_k) ) \} ) +
O(\delta),
\end{eqnarray*}
uniformly over $t_m\leq s_{k+1}\leq t_{m+1}, 0\leq s_1 \leq \cdots \leq
s_k\leq t_m$. Consequently, (\ref{eq:a.20}) equals
\begin{eqnarray*}
&& \sum_{k= 0}^{2 \lceil T/\theta\rceil}  \sum_{i_1,\ldots, i_k\in
\{1,2\}} \lim_{\delta \rightarrow 0} \Biggl\{ \int_0^{t_m} \ldots
\int_{s_{k-1}}^{t_m} \lambda (t_m, (1)| \{ (s_1, (i_1) ), \ldots, (s_k,
(i_k) ) \} )\\
&&\hphantom{\sum_{k= 0}^{2 \lceil T/\theta\rceil}  \sum_{i_1,\ldots, i_k\in
\{1,2\}} \lim_{\delta \rightarrow 0} \Biggl\{}{}\times \Biggl[ \prod_{l=1}^k \lambda (s_l, (i_l) | \{ (s_1,
(i_1) ), \ldots, (s_{l-1}, (i_{l-1}) ) \} ) \Biggr]\\
&& \hphantom{\sum_{k= 0}^{2 \lceil T/\theta\rceil}  \sum_{i_1,\ldots, i_k\in
\{1,2\}} \lim_{\delta \rightarrow 0} \Biggl\{}{}\times e^{-\sum_{j=1}^2 \int_0^{t_m} \lambda (w, (j) |
{\mathcal H}_w)\,dw}\,ds_{k+1}\,ds_k\,\cdots\,ds_1 + O(\delta)\Biggr\}\\
&&\qquad= E[ \lambda (t, (1) |{\mathcal H}_t],
\end{eqnarray*}
since $t_{m+1}-t_m =\delta$ and $\lim_{\delta\rightarrow 0} t_m= t$.
Using a similar argument, we have
\begin{eqnarray*}\label{eq:2.5}
\frac{ P^{1, 2}_{1, 1} (t_m) }{\delta^2} &=& \sum_{1\leq \ell_1 \neq
\ell_2 \leq 2} \frac{1}{\delta^2} \sum_{k= 0}^{2 \lceil T/\theta
\rceil} \sum_{i_1,\ldots, i_k \in \{1,2\}}\\
&&\int_0^{t_m} \ldots \int_{s_{k-1}}^{t_m} \int_{t_m}^{t_{m+1}} \int_{s_{k+1}}^{t_{m+1}}
\nonumber \\
&&\hspace*{3.3pt}\lambda (s_{k+2}, (2)| \{ (s_1, (i_1) ),
\ldots, (s_k, (i_k) ), (s_{k+1}, (1)) \} )
\nonumber \\
&&{}\times
 \lambda (s_{k+1}, (1)| \{ (s_1, (i_1)), \ldots, (s_k, (i_k)) \} )
\nonumber \\
&& {} \times \Biggl[ \prod_{l=1}^k \lambda (s_l, (i_l)| \{ (s_1,
(i_1)), \ldots, (s_{l-1}, (i_{l-1})) \} ) \Biggr]
\nonumber \\
&& {}\times e^{-\sum_{j=1}^2 \int_0^{t_{m+1}} \lambda (w,
(j)| {\mathcal H}_w)\,dw}\,ds_{k+2}\,ds_{k+1}\,ds_k\,\cdots\,ds_1
\nonumber \\
&\rightarrow & \frac{1}{2} \bigl\{ E \bigl[ \lambda \bigl(t, (2) | \{(t,
(1)) \} \cup {\mathcal H}_t \bigr) \lambda (t, (1)| {\mathcal H}_t)  \bigr]\\
&&\hphantom{\frac{1}{2} \bigl\{}{} + E \bigl[ \lambda \bigl(t, (1) | \{(t, (2))\} \cup {\mathcal H}_t \bigr)
\lambda (t, (2)| {\mathcal H}_t) \bigr] \bigr\}
\end{eqnarray*}
and $\delta^{-1} P^1_1 (t_m) = \delta^{-1} [ P^{1,2}_{1,0} (t_m) +
P^{1,2}_{1,1} (t_m)] \rightarrow E[ \lambda (t, (1) |{\mathcal H}_t]$ as
$\delta \rightarrow 0$.
\end{pf*}

\begin{pf*}{Proof of Theorem \ref{tm:4.2}} For simplicity, define
\[
P^{1,2;1,2}_{a,b;c,d} (t_m, t_{m+h}) = P \bigl( X^1( t_m)=a, X^2(t_m)=b,
X^1(t_{m+h}) =c, X^2(t_{m+h}) =d \bigr)
\]
for all $a,b,c,d,\in \{0,1\}$. We observe from Proposition \ref{pn:8.1}
that
\begin{eqnarray*}
&&\hspace*{-4pt} \delta^{-2} P^{1, 2; 1,2}_{1, 0; 0; 1} (t_m, t_{m+h}) \\
&&\hspace*{-4pt}\qquad=  \frac{1}{\delta^2} \sum_{j=0}^{2 \lceil T/\delta\rceil} \sum_{k =
0}^{2 \lceil T/\theta \rceil}
 \sum_{i_1,\ldots, i_j, i_{j+2},\ldots,  i_{j+k+1} \in \{1,2\}}\\
 &&\qquad\quad\int_0^{t_m} \ldots \int_{s_{j-1}}^{t_m} \int_{t_m}^{t_{m+1}} \int_{s_{j+1}}^{t_{m+h}}
\cdots \int_{s_{j+ k}}^{t_{m+h}} \int_{ t_{m+h}}^{t_{m+h+1}} \\
&&\hspace*{-4pt}\qquad\quad\hspace*{3.3pt}\lambda (s_{j+k+2}, (2)| \{ (s_1,
(i_1) ), \ldots, (s_j, (i_j) ), (s_{j+1}, (1)), (s_{j+2}, (i_{j+2})),\\
&&\hspace*{-4pt}\qquad\quad\hspace*{189pt}\ldots, (s_{j+k+1}, (i_{j+k+1})) \} )
\nonumber \\
&&\hspace*{-4pt}\qquad \quad{}\times \Biggl[ \prod_{l= j+2}^{j+k+1} \lambda (s_l,
(i_l)| \{ (s_1, (i_1)), \ldots, (s_j, (i_j)),\\
&&\hspace*{-4pt}\qquad\quad\hspace*{46pt} (s_{j+1}, (1)),
(s_{j+2}, (i_{j+2})),\ldots, (s_{l-1}, (i_{l-1})) \} ) \Biggr]\\
&&\hspace*{-4pt}\qquad\quad{}\times \lambda
(s_{j+1}, (1)| \{ (s_1, (i_1) ), \ldots, (s_j, (i_j) )\} )
\nonumber \\
&&\hspace*{-4pt}\qquad\quad{} \times \Biggl[ \prod_{l= 1}^j \lambda (s_l, (i_l)|
\{ (s_1, (i_1)), \ldots, (s_{l-1}, (i_{l-1})) \} ) \Biggr]
\nonumber \\
&&\hspace*{-4pt}\qquad\quad{} \times e^{-\sum_{\ell =1}^2 \int_0^{t_{m+1}} \lambda
(w, (\ell )| {\mathcal
H}_w)\,dw}\,ds_{j+k+2}\,ds_{j+k+1}\,\cdots\,ds_{j+2}\,
ds_{j+1}\,ds_j\,\cdots\,ds_1 \\
&&\hspace*{-4pt}\qquad\rightarrow  E \bigl[ \lambda \bigl(t+\tau, (2) | \{(t, (1)) \} \cup
{\mathcal H}_t \bigr) \lambda (t, (1)| {\mathcal H}_t)  \bigr]
\end{eqnarray*}
as $\delta \rightarrow 0$. Theorem \ref{tm:4.2} follows since
$P^{1,2}_{1,1} (t_m, t_{m+h}) \sim P^{1, 2; 1,2}_{1, 0; 0; 1} (t_m,
t_{m+h})$ as $\delta \rightarrow~0$.
\end{pf*}

\begin{pf*}{Proof of Theorem \ref{tm:5.3}} For simplicity, we only consider
the case $\nu=2$. Let $\bar{\mathcal H}_{t_m}^i = \{ t_{l_{i,1}},\ldots,
t_{l_{i,k_i}}\}$ and $\bigcup_{i=1}^2 \{l_{i,1},\ldots, l_{i,k_i}\} = \{
l_1,\ldots, l_k\}$, where $0\leq l_1 <\cdots < l_k \leq m-1$. We
observe from (\ref{eq:5.6}) that $P^1_1 (t_m | \bar{\mathcal H}_{t_m}) =
P^{1, 2}_{1, 0} (t_m | \bar{\mathcal H}_{t_m}) + O(\delta^2)$, and
\begin{eqnarray*}
\frac{P^{1, 2}_{1, 0} (t_m | \bar{\mathcal H}_{t_m}) P(  \bar{\mathcal
H}_{t_m}) }{\delta}&=& \frac{1+ O(\delta)}{\delta} \int_{t_{l_1}}^{t_{l_1+1}} \ldots
\int_{t_{l_k}}^{t_{l_k+1}} \int_{t_m}^{t_{m+1}}\\
&&\hspace*{1.5pt}\lambda^1 (s_{k+1} | \{
s_{l_{1,1}}, \ldots, s_{l_{1,k_1}} \} )\,
 ds_{k+1}\,d F(s_{l_1}, \ldots, s_{l_k})\\
&=&  \lambda^1 ( t_m | \bar{\mathcal H}^1_{t_m} ) P( \bar{\mathcal H}_{t_m})
+O(\delta) P(  \bar{\mathcal H}_{t_m})
\end{eqnarray*}
for sufficiently small $\delta$ where $F$ denotes the distribution
function of $s_{l_1},\ldots, s_{l_k}$. Here $s_{l_1} \in [t_{l_1},
t_{l_1+1})$, $s_{l_{1,1}} \in [t_{l_{1,1}}, t_{l_{1,1}+1})$, etc. This
proves the first statement of Theorem \ref{tm:5.3}. Next we observe
that
\begin{eqnarray*}
&& \frac{P^{1, 2}_{1, 1} (t_m | \bar{\mathcal H}_{t_m}) P(  \bar{\mathcal
H}_{t_m}) }{\delta^2}\\
&&\qquad= \frac{1+ O(\delta)}{\delta^2 } \int_{t_{l_1}}^{t_{l_1+1}}\ldots
\int_{t_{l_k}}^{t_{l_k+1}}\!\!\! \int_{t_m}^{t_{m+1}} \delta \gamma (s_{k+1})
\lambda^1 (s_{k+1} | \{ s_{l_{1,1}}, \ldots, s_{l_{1,k_1}} \} ) \\
&& \qquad\quad{} \times \lambda^2 (s_{k+1} | \{ s_{l_{2,1}}, \ldots,
s_{l_{2,k_2}} \} )\,ds_{k+1}\,d F(s_{l_1}, \ldots, s_{l_k})\\
&&\qquad\quad{} + \frac{1+ O(\delta)}{\delta^2 } \int_{t_{l_1}}^{t_{l_1+1}} \ldots
\int_{t_{l_k}}^{t_{l_k+1}}\!\!\! \int_{t_m}^{t_{m+1}}\!\!\! \int_{t_m}^{t_{m+1}}
\lambda^1 (s_{k+2} | \{ s_{l_{1,1}}, \ldots, s_{l_{1,k_1}} \} ) \\
&&\qquad\qquad{}\times \lambda^2 (s_{k+1} | \{ s_{l_{2,1}}, \ldots, s_{l_{2,k_2}}
\})\,ds_{k+2}\,ds_{k+1}\,d F(s_{l_1}, \ldots, s_{l_k}) \\
&&\qquad=  [ 1 + \gamma(t_m)] \lambda^1 ( t_m | \bar{\mathcal H}^1_{t_m} )
\lambda^2 ( t_m | \bar{\mathcal H}^2_{t_m} ) P(  \bar{\mathcal H}_{t_m})
+O(\delta) P(  \bar{\mathcal H}_{t_m})
\end{eqnarray*}
for sufficiently small $\delta$. This proves Theorem \ref{tm:5.3}.
\end{pf*}

\begin{pf*} {Proof of Theorem \ref{tm:5.2}} We observe that ${\mathcal H}_{t_m}
\rightarrow \bar{\mathcal H}_{t_m}$ is a many-to-one mapping and from
(\ref{eq:5.4}) that
\begin{eqnarray}\label{eq:a.32}
&& P \bigl( X^1 (t_m)=1, X^2(t_{m+h})=1| {\mathcal H}_{t_m}
\bigr)\nonumber\\
&&\qquad= \delta^2 \gamma (t_m, t_{m+h}) \lambda^1 (t_m| {\mathcal H}^1_{t_m})
\lambda^2 (t_{m+h} | {\mathcal H}^2_{t_{m+h}})\\
&&\qquad\quad{} +  \delta^2 \lambda^1 (t_m|
{\mathcal H}^1_{t_m}) \lambda^2 (t_{m+h} | {\mathcal H}^2_{t_{m+h}}) +
O(\delta^3) \nonumber
\end{eqnarray}
for sufficiently small $\delta$. We further observe that
\begin{eqnarray} \label{eq:a.34}
P^1_1 ( t_m | {\mathcal H}_{t_m}^1 ) &= & P \bigl( X^1 (t_m)=1, X^2(t_{m+h})=0 |
{\mathcal H}_{t_m}^1 \bigr) + O(\delta^2 )\nonumber \\
&=& \delta \lambda^1 (t_m| {\mathcal H}^1_{t_m}) + O(\delta^2)\nonumber\\
&=& \delta \lambda^1 (t_m| \bar{\mathcal H}^1_{t_m}) + O(\delta^2),\nonumber
\\ [-8pt]\\ [-8pt]
P^2_1 ( t_{m+h} | {\mathcal H}_{t_{m+h}}^2 ) &=& P \bigl( X^1 (t_m)=0,
X^2(t_{m+h})=1 | {\mathcal H}_{t_{m+h}}^2 \bigr) + O(\delta^2 )\nonumber \\
&=& \delta  \lambda^2 (t_{m+h} | {\mathcal H}^2_{t_{m+h}} ) + O(\delta^2)\nonumber \\
&=& \delta  \lambda^2 (t_{m+h} | \bar{\mathcal H}^2_{t_{m+h}} ) +O(\delta^2).\nonumber
\end{eqnarray}
 Thus, it follows from (\ref{eq:a.32}) and (\ref{eq:a.34}) that
\begin{eqnarray*}
&& P \bigl( X^1 (t_m)=1, X^2(t_{m+h})=1| \bar{\mathcal H}_{t_m} \bigr)\\
&&\qquad= E \bigl[ P \bigl( X^1 (t_m)=1, X^2(t_{m+h})=1| {\mathcal H}_{t_m} \bigr)
| \bar{\mathcal H}_{t_m} \bigr]\\
&&\qquad= \delta^2 [ \gamma(t_m, t_{m+h}) +1 ] P^1_1 ( t_m | \bar{\mathcal
H}_{t_m}^1 ) P^2_1 ( t_{m+h} | \bar{\mathcal H}_{t_{m+h}}^2 ) +
O(\delta^3)
\end{eqnarray*}
for sufficiently small $\delta$. This proves Theorem
\ref{tm:5.2}.
\end{pf*}
\end{appendix}

\printaddresses

\end{document}